\documentclass[prb,twocolumn,showpacs,amsmath,amssymb,superscriptaddress,floatfix]{revtex4}

\usepackage[english]{babel}
\usepackage{graphicx}

\begin{document}

\title{Multi-impurity Anderson model for quantum dots coupled in parallel}

\author{R. \v{Z}itko}
\affiliation{Jo\v{z}ef Stefan Institute, Ljubljana, Slovenia}

\author{J. \surname{Bon\v ca}}
\affiliation{Faculty of Mathematics and Physics, University of
Ljubljana, Ljubljana, Slovenia}
\affiliation{Jo\v{z}ef Stefan Institute, Ljubljana, Slovenia}

\date{\today}

\begin{abstract}
The system of several ($N$) quantum dots coupled in parallel to the
same single-mode conduction channel can be modelled as a
single-channel $N$-impurity Anderson model. Using the generalized
Schrieffer-Wolff transformation we show that near the particle-hole
symmetric point, the effective Hamiltonian in the local moment regime
is the $N$-impurity $S=1/2$ Kondo model. The conduction-band-mediated
RKKY exchange interaction between the dots is ferromagnetic and at
intermediate temperatures locks the moments into a maximal spin
$S=N/2$ ground state. We provide an analytical estimate for the RKKY
interaction. At low temperatures the spin is partially screened by the
conduction electrons to $N/2-1/2$ due to the Kondo effect. By
comparing accurate numerical renormalization group results for
magnetic susceptibility of the $N-$impuriy Anderson model to the exact
Bethe-Ansatz results of a $S=N/2$ $SU(2)$ Kondo system we show, that at
low-temperature the quantum dots can be described by the effective
$S=N/2$ Kondo model. Moreover, the Kondo temperature is independent of
the number of impurities $N$.  We demonstrate the robustness of the
spin $N/2$ ground state as well as of the associated $S=N/2$ Kondo
effect by studying the stability of the system with respect to various
experimentally relevant perturbations. We finally explore various
quantum phase transitions driven  by these perturbations.
\end{abstract}


\maketitle

\newcommand{\vc}[1]{\boldsymbol{#1}}
\newcommand{\ket}[1]{|#1\rangle}
\newcommand{\bra}[1]{\langle#1|}

\section{Introduction}

The Kondo effect emerges as the increased scattering rate of the
conduction band electrons at low temperatures due to the presence of
magnetic impurities which induce spin-flip scattering. It leads to
various anomalies in the thermodynamic and transport properties of the
Kondo systems. It is usually described using simplified quantum
impurity models such as the Kondo model and the Anderson model
\cite{anderson1961}.The quantum impurity models attract the interest
of the solid state physics community both due to their unexpectedly
complex behavior and intrinsic beauty, as well as due to their
ubiquitous applicability to a vast array of physical systems such as
bulk Kondo systems, heavy-fermion compounds and other strongly
correlated systems \cite{hewson}, dissipative two-level systems
\cite{vladar1983}, single magnetic impurities and quantum dots
\cite{madhavan1998, li1998, cronenwett1998}.

After the properties of single-impurity models were unraveled using a
complementary set of techniques (the scaling approach, Wilson's
numerical renormalisation group, Bethe-Ansatz solution and various
large-$N$ expansion schemes) \cite{hewson}, the attention has
increasingly focused to multi-impurity models. Research in this field
has recently increased due to a multitude  of experimental results
made possible by advances in micro- and nanotechnology. The
multi-impurity magnetic nanostructures under study are predominantly
of two kinds: clusters of magnetic adsorbates on surfaces of noble
metals (Ni dimers \cite{madhavan2002}, Ce trimers \cite{jamneala2001},
molecular complexes \cite{wahl2005}) and systems of multiple quantum
dots \cite{jeong2001, holleitner2002, wiel2003, craig2004, chen2004}.

The most important additional element that emerges in multi-impurity
models is the Ruderman-Kittel-Kasuya-Yosida (RKKY) exchange
interaction \cite{ruderman1954}.  It arises when the magnetic moments
on the impurities induce spin polarization in the conductance band
which leads to magnetic coupling of moments that are separated in
space. The RKKY interaction depends on the inter-impurity distance and
can be either ferromagnetic or antiferromagnetic.

The competition between the antiferromagnetic RKKY interaction and the
Kondo effect in two magnetically coupled local moments leads to a
quantum phase transition at $J \sim T_K$ between strongly bound local
magnetic singlet for $J \gg T_K$ and two separate Kondo singlets for
$J \ll T_K$ \cite{jones1987, jones1988, jones1989, sire1993,
affleck1995}. The role of the antiferromagnetic exchange interaction
was also studied in the context of double quantum dots (DQD)
\cite{georges1999, izumida2000, aono2001, boese2002, lopez2002}. Two
mechanisms can contribute to the effective exchange interaction
between the dots: the conduction-band mediated RKKY interaction and
the super-exchange mechanism due to inter-dot electron
hopping. Depending on the setup (serial or parallel embedding of the
dots between the source and drain leads), either or both mechanisms
may be significant. In magnetically coupled dots, embedded between the
leads in series, the conductance is low for small exchange coupling
when the Kondo singlets are formed between each dot and adjacent
lead. Conductance is also low for large exchange coupling, when a
local singlet state forms between the moments on the dots.  In
contrast, the conductance reaches the unitary limiting value of
$2e^2/h$ in a narrow interval of $J$, such that $J \sim T_K$
\cite{georges1999, izumida2000}. The introduction of additional
electron hopping between dots breaks the quantum critical transition,
nevertheless, some signatures of the quantum phase transition remain
detectable \cite{izumida2000}.

Strong ferromagnetic RKKY interaction between two magnetic impurities
coupled to two conduction channels leads to three different regimes.
At temperatures comparable to RKKY interaction, ferromagnetic locking
of impurity spins occurs; this is followed by a two-stage freezing out
of their local moment as they become screened by the conduction
electrons \cite{jayaprakash1981}. This scenario was corroborated by
numerical studies of the two-impurity Kondo model \cite{silva1996} and
the Alexander-Anderson model \cite{paula1999}.  Antiferromagnetic and
ferromagnetic RKKY interactions lead to different transport properties
of DQD systems \cite{simon2005, vavilov2005}. Due to recent advances
in nanotechnology, the effects of RKKY interaction on transport
properties became directly observable \cite{craig2004}.
Conductance through Aharonov-Bohm (AB) interferometers with embedded
quantum dots also depends on the RKKY interactions, which in turn
depends on the magnetic flux \cite{utsumi2004, lopez2005,
izumida2005}. 
A similar system of two quantum dots, side-coupled to a single-mode channel,
allows to study the crossover between fully screened and underscreened Kondo
impurity \cite{tamura2005}.

The physics of RKKY interactions is also related to the studies of the
Kondo effect in integer-spin quantum dots \cite{sasaki2000}.  By
tuning the magnetic field, the energy difference between singlet and
triplet spin states can be tuned to zero. At the degeneracy point, a
large zero-bias resonance with an increased Kondo temperature is
observed \cite{sasaki2000}, which can be understood in the framework
of a two-orbital Anderson model \cite{izumida2001}.

The interplay of the Kondo effect and the inter-impurity exchange
interaction leads to a number of interesting phenomena observed in
different realizations of the double quantum dot systems.  For this
reason, we present in this work a study of a more general $N$ quantum
dot systems. Using numerical renormalization group (NRG) technique as
our primary tool and various analytical approaches we investigate the
effects of the RKKY interaction in a multi-impurity Anderson model. We
present results of thermodynamic properties, in particular the
impurity contribution to the magnetic susceptibility and the entropy,
as well as various correlation functions. This work also provides a
setting for further studies of transport properties of this class of
systems.

The paper is organized as follows. In section \ref{sec_model} we
describe the class of models under study as well as model parameters
and approximations used in this work.  In section \ref{sec_lowtemp} we
describe the existence of a hierarchy of separated time (and energy)
scales and we introduce effective models valid at different
temperatures. In section \ref{sec_methods} we describe the numerical
methods that are used in section \ref{sec_results} to study the
multi-impurity Anderson models. Finally, in section
\ref{sec_stability} we test the stability of the $S=1$ state in the
two impurity model with respect to various perturbation. Tedious
derivations of scaling equations and perturbation theory approaches
are given in the appendices.

\section{The Model}
\label{sec_model}

\begin{figure}[htbp!]
\centering
\includegraphics[width=6cm,clip]{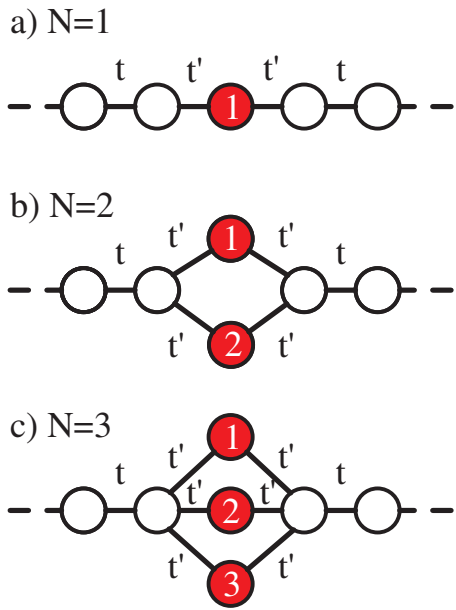}
\caption{(Color online) Systems of parallel quantum dots. The tight-binding hopping
parameter $t$ determines the half-width of the conduction band, $D=2t$,
while parameter $t'$ is related to the hybridization $\Gamma$ by $\Gamma/D
= (t'/t)^2$.}
\label{fig1}
\end{figure}

We study models of $N$ impurities coupled to one single-mode conduction
channel. The motivation for such models comes primarily from experiments
performed on systems of several quantum dots connected in parallel between
source and drain electron reservoirs. Since quantum dots can be made to
behave as single magnetic impurities, such systems can be modelled in the
first approximation as several Anderson impurities embedded between two
tight-binding lattices as shown schematically in Fig.~\ref{fig1}. If the
coupling to the left and right electrode of each quantum dot is symmetric,
it can be shown that each dot couples only to the symmetric combination of
conduction electron wave-functions from left and right lead, while the
antisymmetric combinations of wave-functions are totally decoupled and are
irrelevant for our purpose \cite{glazman1988}. We can thus model the
parallel quantum dots using the following simplified Hamiltonian, which we
name the ``$N$-impurity Anderson model'':
\begin{equation}
\label{Ham}
H = H_{\mathrm{band}} + H_{\mathrm{dots}} + H_{\mathrm{c}}.
\end{equation}
Here $H_{\mathrm{band}} = \sum_{k\sigma} \epsilon_k c^\dag_{k\sigma}
c_{k\sigma}$ is the conduction band Hamiltonian.
$H_{\mathrm{dots}} =  \sum_{i=1}^N H_{\mathrm{dot},i}$ with
\begin{equation}
\begin{split}
H_{\mathrm{dot},i} &= \delta \left( n_i - 1 \right)
+ \frac{U}{2} \left( n_i - 1 \right)^2 \\
&= \epsilon_d n_i + U n_{\uparrow i} n_{\downarrow i}
\end{split}
\end{equation}
is the quantum dot Hamiltonian. Finally,
\begin{equation}
H_{\mathrm{c}} = \frac{1}{\sqrt{L}} \sum_{k\sigma i} \left( V_k d^\dag_{i\sigma} c_{k\sigma} + 
\text{H.c.} \right)
\end{equation}
is the coupling Hamiltonian, where $L$ is a normalization
constant. The number operator $n_i$ is defined as $n_i = \sum_\sigma
d^\dag_{i\sigma} d_{i\sigma}$. Parameter $\delta$ is related to the
more conventional on-site energy $\epsilon_d$ by $\delta =
\epsilon_d+U/2$, where $U$ is the on-site Coulomb electron-electron
(e-e) repulsion. For $\delta=0$ the model is particle-hole symmetric
under the transformation $c^\dag_{k\sigma} \to c_{k,-\sigma}$,
$d^\dag_{i\sigma} \to -d_{i,-\sigma}$. Parameter $\delta$ thus
represents the measure for the departure from the particle-hole
symmetric point.

To cast the model into a form that is more convenient for a numerical
renormalization group study, we make two more approximations. We first
linearize the dispersion relation $\epsilon_k$ of the conduction band,
which gives $\epsilon_k = D k$. The wave-number $k$ runs from $-1$ to
$1$, therefore $2D$ is the width of the conduction band. This
assumption is equivalent to adopting a constant density of states,
$\rho_0 = 1/(2D)$.  Second, we approximate the dot-band coupling with
a constant hybridization strength, $\Gamma = \pi \rho_0
|V_{k_F}|^2$. Neither of these approximations affects the results in a
significant way. In the rest of the paper, we will present results in
terms of the parameters $D$ and $\Gamma$, instead of  the parameters
$t$ and $t'$ of the original tight-binding models depicted in
Fig.~\ref{fig1}.  Our notation follows that of
Refs.~\onlinecite{krishna1980a, krishna1980b} for easier comparison of
the $N$-impurity results with the single-impurity case.

\section{Low-temperature effective models}
\label{sec_lowtemp}

Our primary goal is to demonstrate  that the low-temperature effective model
for the multiple impurity system is the $S=N/2$ $SU(2)$ Kondo model:
\begin{equation}\label{kondo}
H=H_\mathrm{band} + \sum_{k'k} J_{k'k} \vc{s}_{kk'} \cdot \vc{S},
\end{equation}
where $\vc{s}_{kk'} = \frac{1}{2} \sum_{\alpha\alpha'} c^\dag_{k\alpha}
\boldsymbol{\sigma}_{\alpha\alpha'} c_{k'\alpha'}$ is the local-spin density in
the Wannier orbital in the conduction band that couples to {\it all}
$N$ impurities.  $\vc{S}$ is the collective impurity $S=N/2$ spin
operator and $J_{k'k}$ is the momentum-dependent anti-ferromagnetic
spin-exchange interaction that can be derived using the
Schrieffer-Wolff transformation. Results for $J_{k'k}$ are independent
of $N$.

We first argue in favor of the validity of the effective Hamiltonian,
proposed in Eq.~\eqref{kondo}, by considering the different time scales
of the original $N$-impurity Anderson problem. To simplify the
argument  we further focus on the (nearly) symmetric case $\delta \ll
U$ within the Kondo regime, $U/(\Gamma \pi) \gg 1$.

The shortest time scale, $\tau_U \sim \hbar/U$,
represents charge excitations.
The longest time scale is associated with the Kondo effect (magnetic
excitations) and it is given by $\tau_K \sim \hbar / T_K$ where $T_K$
is the Kondo temperature of the {\it single impurity Anderson model},
given by Haldane's expression \label{haldane1978, krishna1980a}
\begin{equation}
T_K = 0.182 U \sqrt{\rho_0 J_\mathrm{K}} \exp \left( - \frac{1}{\rho_0
J_\mathrm{K}} \right),
\label{tkh}
\end{equation}
where $J_K$ is the effective anti-ferromagnetic Kondo exchange
interaction and $\rho_0 J_K=8\Gamma/\pi U$. This expression is
valid for $U \ll D$ and $\delta=0$.

As we will show later, there is an additional time scale $\tau_J \sim
\hbar/J_\mathrm{RKKY}$, originating from the {\it ferromagnetic} RKKY
dot-dot interactions:
\begin{equation}
\label{jeff}
J_{\mathrm{RKKY}} \sim U (\rho_0 J_K)^2 = \frac{64}{\pi^2} \frac{\Gamma^2}{U}.
\end{equation}
From the condition for a well developed Kondo effect, $U/(\Gamma \pi)
\gg 1$, we obtain $J_{\mathrm{RKKY}} \ll U$. We thus establish a
hierarchy of time scales $\tau_U \ll \tau_J \ll \tau_K$.

Based on the three different time-scales, we predict the existence of three
distinct regimes close to the particle-hole symmetric point. The local
moment regime is established at $T \sim T_1^*$, where $T_1^*=U/\alpha$ and
$\alpha$ is a constant of the order one \cite{krishna1980a}.
In this regime the system behaves as $N$ independent spin $S=1/2$
impurities. At $T \sim T_F^*$, where $T_F^* = J_\mathrm{RKKY}/\beta$ and
$\beta$ is a constant of the order one, spins bind into a high-spin $S=N/2$
state. With further lowering of the temperature, at $T\sim T_K$ the $S=N/2$
object experiences the Kondo effect which screens half a unit of spin (since
there is a single conduction channel) to give a ground-state spin of
$S-1/2=(N-1)/2$.

\subsection{Schrieffer-Wolff transformation for multiple impurities}

\newcommand{\Scal}{{\mathcal{S}}}

For $T<T_1^*$, the single impurity Anderson model can be mapped using
the Schrieffer-Wolff transformation \cite{schrieffer1966} to an $s-d$
exchange model (the Kondo model) with an energy dependent
anti-ferromagnetic exchange interaction $J_{k'k}$. In this subsection
we show that for multiple impurities a generalized Schrieffer-Wolff
transformation can be performed and that below $T_1^*$, the
$N$-impurity Anderson model maps to the $N$-impurity $S=1/2$ Kondo
model. Furthermore, the exchange constant is shown to be the same as
in the single impurity case.

Due to the hybridization term $V_k$, the electrons are hopping on
and off the impurities. Since all impurities are coupled to the same
Wannier orbital, it could be expected that these hopping transitions
would somehow ``interfere''. It should be recalled, however, that
the dwelling time $\tau_U$ is much shorter than the magnetic time scales
$\tau_J$ and $\tau_K$.
In other words, spin-flips are realized on
a much shorter time-scale compared to the mean-time between
successive spin-flips; for this reason, each local moment may be
considered as independent. %
Note that the
impurities do in fact ``interfere'': there are ${\mathcal O}(V_k^4)$
processes which lead to an effective ferromagnetic RKKY exchange
interaction between pairs of spins and ultimately to the
ferromagnetic ordering of spins at temperatures below
$\sim J_\mathrm{RKKY}$. This will be discussed in the following
subsection.

The Schrieffer-Wolff transformation is a canonical transformation
that eliminates hybridization terms $V_k$ to first order from the
Hamiltonian $H$, i.e. it requires that \cite{schrieffer1966}
\begin{equation}\label{barh}
{\bar H} \equiv e^\Scal H e^{-\Scal}
\end{equation}
have no terms which are first order in $V_k$. We expand 
${\bar H}$ in terms of nested commutators:
\begin{equation}
{\bar H} = H + [\Scal,H] + \frac{1}{2} [\Scal,[\Scal,H]] + \ldots
\end{equation}
and  write $H=H_0+H_c$, where $H_0 = H_\mathrm{band} + 
H_\mathrm{dots}$.  We then choose $\Scal$ to be first order in $V_k$ so
that
\begin{equation}
[\Scal,H_0]+H_c=0.
\end{equation}
As previously discussed, each impurity can be considered independent
due to the separation of time scales. Therefore, we choose the
generator $\Scal$ to be the sum $\Scal=\sum_i \Scal_i$ of generators $\Scal_i$,
where the generator $\Scal_i$ for each impurity has the same form as in
the single-impurity case:
\begin{equation}
\label{Si}
\Scal_i = \sum_{k\sigma\alpha} \frac{V_k}{\epsilon_k-\epsilon_\alpha} n_{i,-\sigma}^\alpha
c_{k\sigma}^\dag d_{i\sigma} - \text{H.c.}
\end{equation}
with $\epsilon_{\pm} = \delta \pm U/2$ and the projection operators
$n_{i,-\sigma}^\alpha$ are defined by
\begin{equation}
\begin{split}
n_{i,-\sigma}^+ &= n_{i,-\sigma}, \\
n_{i,-\sigma}^- &= 1-n_{i,-\sigma}.
\end{split}
\end{equation}
The resulting effective Hamiltonian is then given by
\begin{equation}\label{heff}
H_\mathrm{eff} = H_0 + \frac{1}{2} [\Scal,H_c],
\end{equation}
which features $\mathcal{O}(V_k^2)$ effective interactions with the
leading terms that can be cast in the form of the Kondo
antiferromagnetic exchange interaction
\begin{equation}\label{hex}
H_{\mathrm{ex}} = \sum_i \left( \sum_{kk'} J_{k'k} \vc{s}_{kk'}
\cdot \vc{S}_i \right),
\end{equation}
where $\vc{S}_i$ is the $S=1/2$ spin operator on impurity $i$ defined by
$\vc{S}_i = \frac{1}{2} \sum_{\alpha\alpha'} d^\dag_{i\alpha}
\boldsymbol{\sigma} d_{i\alpha'}$ and the exchange constant $J_{k'k}$ is
given by
\begin{equation}
\begin{split}
J_{k'k} = & V_{k'} V_k
\Bigl(
\frac{1}{\epsilon_k - (\delta + U/2)}
+
\frac{1}{\epsilon_{k'} - (\delta + U/2)} \\
& -
\frac{1}{\epsilon_k - (\delta - U/2)}
-
\frac{1}{\epsilon_{k'} - (\delta - U/2)}
\Bigr).
\end{split}
\end{equation}
If we limit the wave-vectors to the Fermi surface, {\it i.e.} for
$k=k'=k_F$, we obtain
\begin{equation}
\label{JKdelta}
J_K \equiv 2|V_{k_F}|^2
\left(
\frac{1}{|\delta-U/2|}+\frac{1}{|\delta+U/2|}
\right).
\end{equation}
This result is identical to $J_{k'k}$ obtained for a single
impurity \cite{schrieffer1966}.

As it turns out, the Schrieffer-Wolff transformation,
Eqs.~\eqref{barh}-\eqref{heff}, produces inter-impurity interaction terms in
addition to the expected impurity-band interaction terms. In the
particle-hole symmetric case ($\delta=0$), these additional terms can be
written as
\begin{equation}
\label{DH2} \Delta H_\mathrm{eff} = 2 \frac{|V_k|^2}{U} \left(
\sum_{i=1}^N n_i - N \right) h_{\mathrm{hop}},
\end{equation}
where
\begin{equation}
h_\mathrm{hop} = \sum_{i<j,\sigma} \left( d^\dag_{i\sigma}
d_{j\sigma} + d^\dag_{j\sigma} d_{i\sigma} \right).
\end{equation}
Since the on-site charge repulsion favors states with single occupancy of
each impurity, the term in the parenthesis in Eq.~\eqref{DH2} is on the
average equal to zero. Furthermore, if each site is singly occupied,
possessing small fluctuations of the charge $\langle n_i^2\rangle-\langle
n_i\rangle^2 \sim 0$, hopping between the sites is suppressed and the term
$h_{\mathrm{hop}}$ represents another small factor.  The Hamiltonian $\Delta
H$ is thus clearly not relevant: impurities are indeed independent.

On departure from the particle-hole symmetric point ($\delta\neq
0$), $\Delta H_{\mathrm{eff}}$ generalizes to
\begin{equation}
\Delta H_{\mathrm{eff}} = 2 \frac{U |V_k|^2}{U^2-4\delta^2} \left( \left(
\sum_{i=1}^2 n_i - N \right) -2 N \frac{\delta}{U} \right)
h_{\mathrm{hop}}.
\end{equation}
For moderately large $\delta/U$ this Hamiltonian term still represents only
a small correction to Eq.~\eqref{hex}. However, for strong departure from
the particle-hole (p-h) symmetric point, close to the valence-fluctuation
regime (i.e. $\delta \to U/2$), the $\Delta H_{\mathrm{eff}}$ becomes
comparable in magnitude to $H_{\mathrm{ex}}$ and generates hopping of
electrons between the impurities.

The above discussion leads us to the conclusion that just below
$T_1^*$ the effective Hamiltonian close to the p-h symmetric point  is
\begin{equation}
\label{kondoeff}
H_\mathrm{eff} = H_\mathrm{band} + \sum_i
\sum_{k'k} J_{k'k} \vc{s}_{kk'} \cdot \vc{S}_i.
\end{equation}
If the dots are described by unequal Hamiltonians $H_{\mathrm{dot},i}$
or have unequal hybridizations $V^i_k$, then the mapping of the
multi-impurity Anderson model to a multi-impurity Kondo model still
holds, however with different effective exchange constants
$J_{k'k}^i$.

\subsection{RKKY interaction and ferromagnetic spin ordering}
\label{sec_rkky}

We now show that the effective RKKY exchange interaction between the
spins in the effective $N$-impurity Kondo model, Eq.~\eqref{kondoeff},
is ferromagnetic and also responsible for locking of spins in a state
of high total spin for temperatures below $T<J_{\rm RKKY}$.

The ferromagnetic character of the RKKY interaction is expected, as
shown by the following qualitative argument. We factor out the spin
operators in the effective Hamiltonian Eq.~\eqref{kondoeff}:
\begin{equation}
H_\mathrm{eff} = H_\mathrm{band} + \left( \sum_{k'k} J_{k'k}
\vc{s}_{kk'} \right) \cdot \sum_i \vc{S}_i.
\end{equation}
Spins $\vc{S}_i$ are aligned in the ground state since such
orientation minimizes the energy of the system.
This follows from considering a spin chain with $N$ sites in a
``static magnetic field'' $\sum_{k'k} J_{k'k} \vc{s}_{kk'}$. The
assumption of a static magnetic field is valid due to the separation
of relevant time scales, $\tau_K \gg \tau_J$. States with $S < N/2$
are clearly excited states with one or several ``misaligned'' spins.

Since the inter-dot spin-spin coupling is a special case of the
Ruderman-Kittel-Kasuya-Yosida (RKKY) interaction in bulk systems
\cite{ruderman1954}, a characteristic  functional dependence given by
\begin{equation}
\label{rkkydep}
J_\mathrm{RKKY} \propto U (\rho_0 J_\mathrm{K})^2
= \frac{64}{\pi^2} \frac{\Gamma^2}{U}
= \frac{16 V_{k_F}^4}{U},
\end{equation}
is expected. The factor $U$ in front of $(\rho_0 J_K)^2$ plays the role of a
high-energy cut-off, much like the $0.196 U$ effective-bandwidth factor in
the expression for $T_K$, Eq.~\eqref{tkh}.

Using the Rayleigh-Schr\"odinger perturbation theory we calculated the
singlet and triplet ground state energies $E_S$ and $E_T$ to the fourth
order in $V_k$ for the two-impurity case (see Appendix~\ref{app_rkky}). We
define the RKKY exchange parameter by $J_\mathrm{RKKY} = E_S - E_T$;
positive value of $J_\mathrm{RKKY}$ corresponds to ferromagnetic RKKY
interaction. For $U/d \lesssim 0.1$, the prefactor of $(\rho_0 J_K)^2$ in
the expression \eqref{rkkydep} is indeed found to be linear in $U$. Together
with the prefactor the perturbation theory leads to
\begin{equation}
\label{ptrkky}
J_\mathrm{RKKY} = 0.62 U (\rho_0 J_K)^2 \quad \text{for}\quad U/D \ll 1,
\end{equation}
which, as we will show later, fits very well our numerical results. The RKKY
interaction becomes fully established for temperatures below $T_J$ which is
roughly one or two orders of magnitude smaller than $T_1^*$ ($T_J$ is
defined in Appendix~\ref{app_rkky}). Since the RKKY interactions in the
first approximation do not depend much on the number of impurities, for $N >
2$ the exchange interaction between each pair of impurities has the same
strength as in the two impurity case. Therefore, for temperatures just below
$T_J$, the effective Hamiltonian for the $N$-impurity Anderson model becomes
\begin{equation}
H_\mathrm{eff} = H_\mathrm{band} +
\left( \sum_{k'k} J_{k'k} \vc{s}_{kk'} \right)
\cdot
\sum_i \vc{S}_i - J_{\mathrm{RKKY}}
\sum_{i < j} \vc{S}_i \cdot \vc{S}_j.
\end{equation}
When the temperature drops below a certain temperature $T_F^*$, the spins
align and form a ferromagnetically-frozen state of maximum spin $S=N/2$. The
transition temperature $T_F^*$ is generally of the same order as
$J_\mathrm{RKKY}$, i.e. $T_F^* = J_\mathrm{RKKY}/\beta$, where $\beta$ is an
$N$-dependent constant of the order one. This relation holds if $T_F^* \ll
T_J$, otherwise $T_F^*$ needs to be determined using a self-consistency
equation~\eqref{selfconsist}, as discussed in Appendix~\ref{app_rkky}.

In conclusion, for $T \lesssim T_F^*$ the states with total spin less than
$N/2$ can be neglected, and the system behaves as if it consisted of a
single spin $\vc{S}$ of magnitude $S=N/2$. The effective Hamiltonian at very
low temperatures is therefore the $S=N/2$ $SU(2)$ Kondo model
\begin{equation}
H_\mathrm{eff} = H_\mathrm{band} + \sum_{k,k'} J_{k'k} \vc{s}_{k,k'} \cdot
\vc{S},
\end{equation}
where $\vc{S}={\mathcal P} \left( \sum_i \vc{S}_i \right) {\mathcal
P}$ and ${\mathcal P}$ is the projection operator on the subspace with
total spin $S=N/2$.  Other multiplets are irrelevant at temperatures
below $T_F^*$. We point out that the Kondo temperature for this model
is given by the formula for the single impurity Anderson model,
Eq.~\eqref{tkh}, irrespective of the number of dots $N$, since the
ferromagnetic interaction only leads to moment ordering, while the
exchange interaction of the collective spin is still given by the same
$J_{k'k}$.

It should be mentioned that if the exchange constants $J^i_{k'k}$ for
different impurities are different, there will be some mixing between the spin
multiplets. The simple description of impurities as a collective $S=N/2$
spin still holds even for relatively large differences, but in general the
virtual excitations to other spin multiplets must be taken into account.
This is studied in detail for the case of two dots in Section~\ref{unequal}.

\section{The methods}
\label{sec_methods}

\subsection{Numerical renormalization group}

The method of choice to study the low-temperature properties of
quantum impurity models is the Wilson's numerical renormalization
group (NRG) \cite{wilson1975, krishna1980a, krishna1980b}. The NRG
technique consists of logarithmic discretization of the conduction
band described by $H_\mathrm{band}$, mapping onto a one-dimensional
chain with exponentially decreasing hopping constants, and
iterative diagonalization of the resulting Hamiltonian. Since all
$N$ impurities couple to the band in the same manner, they
all couple to the same, zero-th site of the chain Hamiltonian
\cite{krishna1980a}:
\begin{equation}
\begin{split}
\frac{H_C}{D} &= \frac{1}{2} (1+\Lambda^{-1}) \\
&\sum_{n=0}^\infty
\sum_\sigma \Lambda^{-n/2} \xi_n \left[ f^\dag_{n,\sigma} f_{n+1,\sigma}
+f^\dag_{n+1,\sigma} f_{n,\sigma} \right] \\
&+ H_{\mathrm{dots}}
+ \sum_{i,\sigma} \left( \frac{2\Gamma}{\pi D} \right)^{1/2}
\left( f^\dag_{0\sigma} d_{i\sigma} + d^\dag_{i\sigma} f_{0\sigma} \right).
\end{split}
\end{equation}
Here $f^\dag_{n\sigma}$ are the chain creation operators and $\xi_n$ are
constants of order 1. In addition to the conventional Wilson's
discretization scheme \cite{wilson1975}, we also used Campo and Oliveira's
new discretization approach using an over-complete basis of states
\cite{campo2005} with $\Lambda=4$, which improved convergence to the
continuum limit. We made use of the ``$z$-trick'' with typically 6 equally
spaced values of the parameter $z$ \cite{oliveira1994}.

\subsubsection{Symmetries}

The Hamiltonian \eqref{Ham} has the following symmetries: a)
$U(1)_\mathrm{gauge}$ symmetry due to global phase (gauge)
invariance. The corresponding conserved quantity is the total charge
(defined with respect to half-filling case): $Q = \sum_i \left( n_i -
1 \right)$, where the sum runs over all the impurity as well as the
lead sites; b) $SU(2)_\mathrm{spin}$ spin symmetry with generators
$\vc{S} = \sum_i \frac{1}{2} \sum_{\alpha\alpha'} a^\dag_{i\alpha}
\boldsymbol{\sigma}_{\alpha\alpha'} a_{i\alpha'}$, where
$\boldsymbol{\sigma}$ are the Pauli matrices. Since operators $Q$,
$\vc{S}^2$ and $S_z$ commute with $H$, the invariant subspaces can be
classified according to quantum numbers $Q$, $S$ and
$S_z$. Computation of matrix elements can be further simplified using
the Wigner-Eckart theorem \cite{krishna1980a}.

In the particle-hole symmetric point, {\it i.e.} $\delta=0$,
Hamiltonian has an additional $SU(2)_\mathrm{iso}$ isospin symmetry
\cite{jones1988}. We define isospin operators on impurity site $i$
using
\begin{equation}
\vc{I}_i = \sum_{\alpha\alpha'}\eta^\dag_{i,\alpha}
\boldsymbol{\sigma}_{\alpha\alpha'} \eta_{i,\alpha'},
\end{equation}
where the Nambu spinor $\eta^\dag_i$ on the impurity orbitals is defined by
\begin{equation}
\eta^\dag_i = \begin{pmatrix}
d^\dag_{i,\uparrow} \\
-d_{i,\downarrow}
\end{pmatrix}.
\end{equation}
We also define $I^+ = I^x + i I^y$ and $I^- = (I^+)^\dag$. We then have, for
example, $I^z_i = (n_i - 1)/2 = Q_i/2$ and $I^+_i = d^\dag_{i\downarrow}
d^\dag_{i\uparrow}$. The isospin symmetry is thus related to the electron
pairing. In terms of the isospin operators the impurity Hamiltonian can be
written as
\begin{equation}
H_{\mathrm{dot},i} = 2\delta I^z_i + 4 U (I^z_i)^2 
= 2\delta I^z_i + \frac{4}{3} U (I_i)^2,
\end{equation}
where we took into account that for spin-1/2 operators (Pauli
matrices) $(I_i^z)^2=1/3 (I_i)^2$.

On the Wilson chain the isospin is defined similarly but with
a sign alternation in the definition of the Nambu spinors $\xi_n$:
\begin{equation}
\xi^\dag_n = \begin{pmatrix}
f^\dag_{n,\uparrow} \\
(-1)^n f_{n,\downarrow}
\end{pmatrix}.
\end{equation}

The total isospin operator is obtained through a sum of $\vc{I}_i$ for
all orbitals of the problem (impurities and conduction band). For
$\delta=0$, both $\vc{I}^2$ and $I_z$ commute with $H$ and $I$ and
$I_z$ are additional good quantum numbers. Note that $I_z = Q/2$,
therefore $U(1)_\mathrm{gauge}$ is in fact a subgroup of
$SU(2)_\mathrm{iso}$. Due to isotropy in isospin space, the $I_z$
dependence can again be taken into account using the Wignert-Eckart
theorem.

Spin and isospin operators commute, $[S_i, I_j]=0$ for all $i,j$.
Therefore, for $\delta=0$ the problem has a $SU(2)_\mathrm{spin}
\otimes SU(2)_\mathrm{iso}$ symmetry which, when explicitly taken
into account, leads to a further significant reduction of the
numerical task.

In all our NRG calculations we took into account the conservation of the
charge and the rotational invariance in the spin space, i.e.  the
$U(1)_\mathrm{gauge} \otimes SU(2)_\mathrm{spin}$ symmetry which holds for
all perturbed models considered, or the $SU(2)_\mathrm{spin} \otimes
SU(2)_\mathrm{isospin}$ symmetry where applicable. The number of states that
we kept in each stage of the NRG iteration depended on the number of the
dots $N$, since the degeneracy increases exponentially with $N$:
approximately as $4^N$ at the high-temperature free orbital regime and as
$2^N$ in the local-moment regime. In the most demanding $N=4$ calculation we
kept up to 12000 states at each iteration (which corresponds to $>32000$
states taking into account the spin multiplicity of states), which gave
fully converged results for the magnetic susceptibility.

For large scale NRG calculations it is worth taking into account that
the calculation of eigenvalues scales as ${\mathcal O}(n^2)$ and the
calculation of eigenvectors as ${\mathcal O}(n^3)$, where $n$ is the
dimension of the matrix being diagonalized. Since eigenvectors of the
states that are truncated are not required to recalculate various
matrices prior to performing a new iteration, considerable amount of
time can be saved by not calculating them at all.

\subsubsection{Calculated quantities}

We have computed the following thermodynamic quantities
\begin{itemize}

\item the temperature-dependent impurity contribution to the magnetic
susceptibility $\chi_\mathrm{imp}(T)$
\begin{equation}
\chi_\mathrm{imp}(T) = \frac{(g\mu_B)^2}{k_B T}
\left(
\langle S_z^2 \rangle
-
\langle S_z^2 \rangle_0
\right)
\end{equation}
where the subscript $0$ refers to the situation when no impurities are
present (i.e. $H$ is simply the band Hamiltonian $H_\mathrm{band}$),
$g$ is the electronic $g$ factor, $\mu_B$ the Bohr magneton and $k_B$
the Boltzmann's constant.  It should be noted that the combination $T
\chi_\mathrm{imp} / (g\mu_B)^2$ can be considered as an effective
moment of the impurities, $\mu_\mathrm{eff}$.

\item the temperature-dependent impurity contribution to the entropy
$S_\mathrm{imp}(T)$
\begin{equation}
S_\mathrm{imp}(T) = \frac{\left( E-F \right)}{T}
- \frac{\left( E-F \right)_0}{T},
\end{equation}
where $E = \langle H \rangle = \mathrm{Tr} \left( H e^{-H/(k_B T)}
\right)$ and $F = -k_B T \ln \mathrm{Tr} \left( e^{-H/k_B T} \right)$.
From the quantity $S_\mathrm{imp}/k_B$ we can deduce the effective
degrees-of-freedom $\nu$ of the impurity as $S_\mathrm{imp}/k_B \sim
\ln \nu$.

\item thermodynamic expectation values of various operators
such as the on-site occupancy $\langle n_i \rangle$, local
charge-fluctuations $\langle (\delta n)^2 \rangle = \langle n_i^2 \rangle -
\langle n_i \rangle^2$, local-spin $\langle \vc{S}_i^2 \rangle$ and
spin-spin correlations $\langle \vc{S}_i \cdot \vc{S}_j \rangle$.
\end{itemize}

In the following we drop the suffix $\mathrm{imp}$ in $\chi_\mathrm{imp}$, 
but one should keep in mind that impurity contribution
to the quantity is always implied. We also set $k_B=1$.

\subsection{Bethe Ansatz}

The single-channel $SU(2)$ Kondo model can be exactly solved for an
arbitrary spin of the impurity using the Bethe Ansatz (BA) method
\cite{andrei1983, rajan1982, sacramento1989}. This technique gives
exact results for thermodynamic quantities, such as magnetic
susceptibility, entropy and heat capacity. It is, however, incapable
of providing spectral and transport properties.  For the purpose of
comparing results of the single-channel $SU(2)$ Kondo model with NRG
results of the N-impurity Anderson model, we have numerically solved
the system of coupled integral equations using a discretization scheme
as described, for example, in Ref.~\onlinecite{sacramento1989}.

\subsection{Scaling analysis}

Certain aspects of the Kondo physics can be correctly captured using the
perturbative renormalization group approach based on the ``poor-man's
scaling'' technique due to Anderson \cite{anderson1970}. A brief account
of this method is given in Appendix~\ref{scaling}.

\section{Numerical results}
\label{sec_results}

We choose the parameters $U$ and $\Gamma$ so that the relevant energy
scales are well separated which enables clear identification of
various regimes and facilitates analytical predictions.

\begin{figure}[htbp!]
\includegraphics[width=8cm,clip]{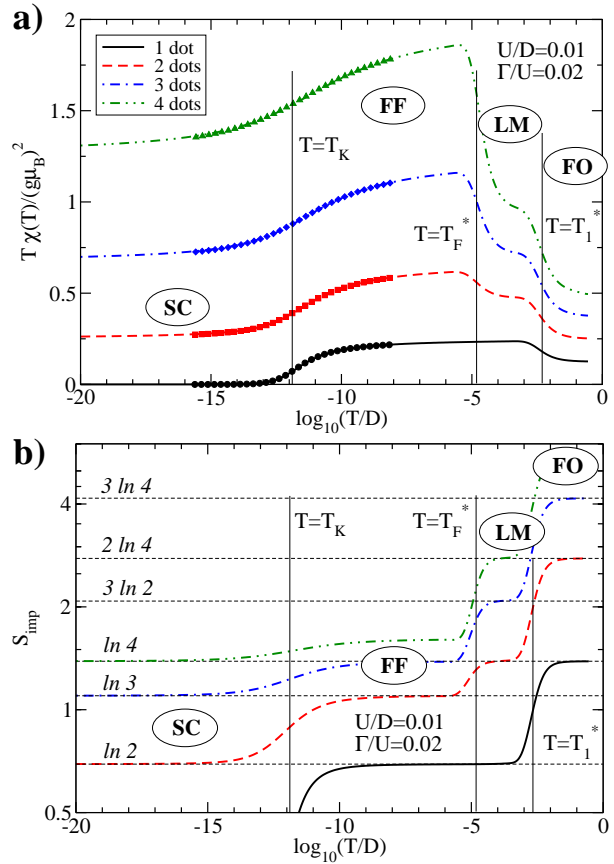}
\caption{(Color online) a) Temperature-dependent susceptibility 
and b) entropy of the $N$-dot systems calculated using the NRG. 
The symbols in the susceptibility
plots were calculated using the thermodynamic Bethe Ansatz approach for the 
corresponding $S=N/2$ SU(2) Kondo models ($\bullet$ $S=1/2$, $\blacksquare$
$S=1$, $\blacklozenge$ $S=3/2$, $\blacktriangle$ $S=2$).}
\label{fig_aa}
\end{figure}

In Fig.~\ref{fig_aa} we show temperature dependence of magnetic
susceptibility and entropy for $N=1,2,3$ and 4 systems. As the
temperature is reduced, the system goes through the following
regimes:
\begin{enumerate}
\item At high temperatures, $T>T_1^*$, the impurities are independent
and they are in the {\it free orbital regime} (FO) (states $\ket{0}$,
$\ket{\uparrow}$, $\ket{\downarrow}$ and $\ket{2}$ on each impurity
are equiprobable). Each dot then contributes $1/8$ to
$\mu_\mathrm{eff}=T\chi/(g\mu_B)^2$ for a total of
$\mu_\mathrm{eff}=N/8$. The entropy approaches $S_\mathrm{imp}=N \ln 4$ since
all possible states are equally probable \cite{krishna1980a}.

\item For $T_F^*<T<T_1^*$ each dot is in the {\it local-moment
regime} (LM) (states $\ket{\uparrow}$ and $\ket{\downarrow}$ are
equiprobable, while the states $\ket{0}$ and $\ket{2}$ are
suppressed). Each dot then contributes $1/4$ to $\mu_\mathrm{eff}$ for
a total of $N/4$. The entropy decreases to $S_\mathrm{imp}=N \ln 2$.

\item For $T_K<T<T_F^*$ and $N\geq 1$ the dots lock into a
high spin state $S=N/2$ due to ferromagnetic RKKY coupling between
local moments formed on the impurities. This is the {\it
ferromagnetically frozen regime} (FF) \cite{jayaprakash1981} with
$\mu_\mathrm{eff}=S(S+1)/3=N/2(N/2+1)/3$. The entropy decreases
further to $S_\mathrm{imp} = \ln(2S+1) = \ln (N+1)$.

\item Finally, for $T<T_K$, the total spin is screened from $S=N/2$ to
$\tilde S = S-1/2 = (N-1)/2$ as we enter the partially-quenched, Kondo
screened {\it strong-coupling} (SC) $N$-impurity regime with
$\mu_\mathrm{eff} = \tilde S(\tilde S+1)/3 =
(N-1)/2[(N-1)/2+1]/3$. The remaining $S-1/2$ spin is a complicated
object: a $S=N/2$ multiplet combination of the impurity spins
antiferromagnetically coupled by a spin-$1/2$ cloud of the lead
\cite{jayaprakash1981}. In this regime, the entropy reaches its
minimum value of $S_\mathrm{imp} = \ln (2{\tilde S}+1) = \ln N$.
\end{enumerate}

\begin{table}[htbp!]
\centering
\begin{ruledtabular}
\begin{tabular}{@{}ccc@{}}
$N$ & Kondo temperature $T_K/D$ & LM-FO temperature $T_F^*/D$ \\
\colrule
1 & $1.20 \times 10^{-12}$ & - \\
2 & $1.23 \times 10^{-12}$ & $1.87 \times 10^{-5}$\\
3 & $1.29 \times 10^{-12}$ & $2.11 \times 10^{-5}$\\
4 & $1.32 \times 10^{-12}$ & $2.32 \times 10^{-5}$\\
\end{tabular}
\end{ruledtabular}
\caption{Kondo temperatures for different numbers of quantum dots $N$
corresponding to plots in Fig.~\ref{fig_aa}. \label{tab_aa}}
\end{table}

In Fig.~\ref{fig_aa}, atop the NRG results we additionally plot the
results for the magnetic susceptibility of the $S=N/2$ $SU(2)$ Kondo
model obtained using an exact thermodynamic Bethe-Ansatz method. For
$T<T_F^*$ nearly perfect agreement between the $N$-impurity Anderson
model and the corresponding $S=N/2$ $SU(2)$ Kondo model are found over
many orders of magnitude. \footnote{Similar comparisons of NRG and BA
results were used in studying the two-stage Kondo screening in the
two-impurity Kondo model (Ref.~\onlinecite{silva1996}) to identify
the nature of the first and the second Kondo cross-over.} This
agreement is used to extract the Kondo temperature of the
multiple-impurity Anderson model. The fitting is performed numerically
by the method of least-squares; in this manner very high accuracy of
the extracted Kondo temperature can be achieved.  The results in
Table~\ref{tab_aa} point out the important result of this work that
the Kondo temperature is nearly independent of $N$, as predicted in
Section~\ref{sec_rkky}. In this sense, the locking of spins into a
high-spin state does not, by itself, weaken the Kondo effect
\cite{craig2004, vavilov2005}; however, it does modify the
temperature-dependence of the thermodynamic and transport properties
\cite{mehta2005, koller2005}.

\begin{figure}[htbp!]
\includegraphics[width=8cm,clip]{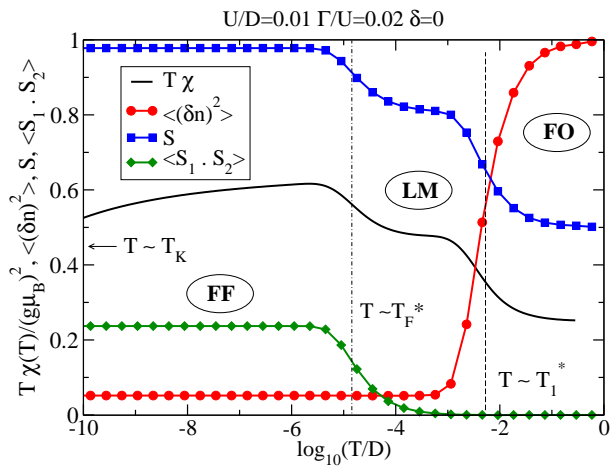}
\caption{(Color online) Temperature-dependence of susceptibility,
charge fluctuations $\langle (\delta n)^2 \rangle$,
total spin $S$ and the spin-spin correlations 
$\langle \vc{S}_1 \cdot \vc{S}_2 \rangle$
of the 2-dot system.}
\label{fig_g}
\end{figure}

It is instructive to follow transitions from high-temperature FO
regime to LM and FF regime through a plot combining the temperature
dependence of the magnetic susceptibility and of other thermodynamic
quantities, as presented in Fig.~\ref{fig_g}.  Charge fluctuations
$\langle (\delta n)^2\rangle$ show a sudden drop at $T \sim T_1^*$
representing the FO - LM transition. In contrast, the magnitude of the
total spin $S$ increases in steps: $S=1/2$, $(\sqrt{7}-1)/2$ and
$1$. Values of $S$ in these plateaus are the characteristic values of
doubly occupied double-quantum dot system in the FO, LM and FF regime,
respectively.

The LM-FF transition temperature $T_F^*$ can be deduced from
the temperature dependence of the spin-spin correlation function. In the FF
regime the spins tend to align, which leads to $\langle {\bf S}_1 \cdot {\bf
S}_2 \rangle \to \sim 1/4$ as $T \to 0$, see Fig.~\ref{fig_g}. The
transition from 0 to $1/4$ is realized at $T \sim T_F^*$. We can extract
$T_F^*$ using the (somewhat arbitrary) condition
\begin{equation}
\label{extract}
\langle {\bf S}_1 \cdot {\bf S}_2 \rangle (T_F^*) = 1/2 \langle
{\bf S}_1 \cdot {\bf S}_2 \rangle (T \to 0).
\end{equation}
In section~\ref{IVc} we show that this condition is in very good
agreement with $T_F^*=J_\mathrm{RKKY}/\beta$ obtained by determining
the explicit inter-impurity antiferromagnetic coupling constant
$J_{12}$, defined by the relation $J_\mathrm{RKKY}+J_{12}=0$ that
destabilizes the high-spin $S=N/2$ state. The extracted $T_F^*$
transition temperatures that correspond to plots in Fig.~\ref{fig_aa}
are given in Table~\ref{tab_aa}. 
We find that they weakly depend on
the number of impurities, more so than the Kondo temperature. The
increase of $T_F^*$ with $N$ can be partially explained by calculating
$T_F^*$ for a spin Hamiltonian $H = -J_\mathrm{RKKY} \sum_{i<j}
\vc{S}_i \cdot \vc{S}_j$ for $N$ spins decoupled from leads. Using
Eq.~\eqref{extract} we obtain $T_F^* \approx 1.18\ J_\mathrm{RKKY}$
for $N=2$, $T_F^* \approx 1.36\ J_\mathrm{RKKY}$ for $N=3$ and $T_F^*
\approx 1.55\ J_\mathrm{RKKY}$ for $N=4$.

By performing NRG calculations of $T_F^*$ for other parameters $U$ and
$\Gamma$ and comparing them to the prediction of the perturbation
theory, we found that the simple formula~\eqref{ptrkky} for
$J_{\mathrm{RKKY}}$ agrees very well with numerical results.

\begin{figure}[htbp!]
\includegraphics[width=8cm,clip]{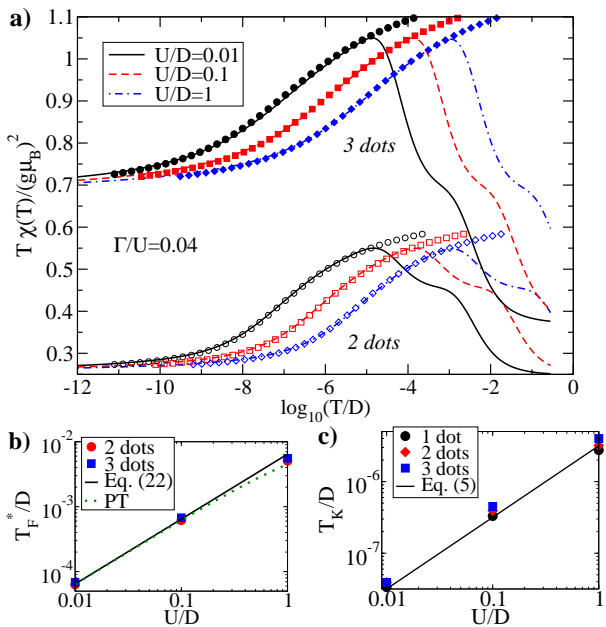}
\caption{(Color online) a) Temperature-dependent susceptibility of the
2 and 3-dot systems
with the same $\Gamma/U$ ratio. Open (filled) symbols are Bethe-Ansatz
results for the $S=1$ ($S=3/2$) Kondo model.
b) Comparison of LM-FF transition temperature $T_F^*$ with predictions
of the perturbation theory. c) Comparison of calculated $T_K$
with the Haldane's formula.
}
\label{fig_a1}
\end{figure}

The effect on thermodynamic properties of varying $U$ while keeping
$\Gamma/U$ (i.e. $\rho_0 J_\mathrm{K}$) fixed is illustrated in
Fig.~\ref{fig_a1} for 2- and 3-dot systems. Parameters $\Gamma$ and $U$
enter expressions for $T_F^*=J_\mathrm{RKKY}/\beta$ and $T_K$ only through
the ratio $\Gamma/U$, apart from the change of the effective bandwidth
proportional to $U$, see Eq.~\eqref{tkh} and \eqref{ptrkky}. This explains
the horizontal shift towards higher temperatures of susceptibility curves
with increasing $U$, as seen in Fig.~\ref{fig_a1}a. The NRG results and the
Bethe-Ansatz for the Kondo models with $S=1$ and $S=3/2$ show excellent
agreement for $T < T_F^*$. In Figs.~\ref{fig_a1}b and \ref{fig_a1}c we
demonstrate the nearly linear $U$-dependence of $T_F^*$ and $T_K$,
respectively.

\begin{figure}[htbp!]
\includegraphics[width=8cm,clip]{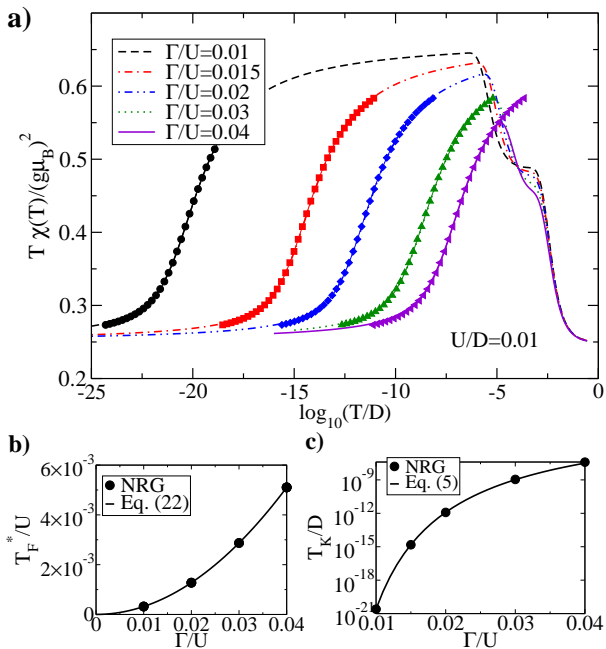}
\caption{(Color online) a) Temperature-dependent susceptibility of the 2-dot system
for equal e-e repulsion $U/D=0.01$ and for different hybridization
strengths $\Gamma$. Symbols represent the Bethe-Ansatz susceptibility for the 
$S=1$ Kondo model with corresponding $T_K$.
b) Comparison of calculated and predicted $T_F^*$.
c) Fit of $T_K$ to the Haldane's formula, Eq.~\eqref{tkh}. 
}
\label{fig_b}
\end{figure}

In Fig.~\ref{fig_b} we show the effect of varying $\Gamma/U$ while keeping
$U$ fixed. In this case, $T_1^*$ stays the same, $T_F^*$ is shifted
quadratically and $T_K$ exponentially with increasing $\Gamma/U$.
Fig.~\ref{fig_b}b shows the agreement of $T_F^*$ with
expression~\eqref{ptrkky}, while Fig.~\ref{fig_b}c shows the agreement of
the extracted values of $T_K$ with formula~\eqref{tkh}.

We note that for $N \geq 2$, eventual coupling to an additional
conduction channel (for example, due to a small asymmetry in the
coupling to the source and drain electrodes) would lead to screening
by additional half a unit of spin \cite{jayaprakash1981, silva1996}
and the residual ground state spin would be $S-1=N/2-1$. For $N \geq
3$ and three channels (due to weak coupling to some third electrode),
three half-units of spin would be screened, and so forth. These
additional stages of Kondo screening would, however, occur at much
lower temperatures; all our findings still apply at temperatures above
subsequent Kondo cross-overs.

In systems of multiple quantum dots, an additional screening mechanism
is possible when after the first Kondo cross-over, the residual
interaction between the remaining spin and the Fermi liquid
quasi-particles is antiferromagnetic \cite{vojta2002}. This leads to
an additional Kondo cross-over at temperatures that are exponentially
smaller than the first Kondo temperature. Such two-stage Kondo effect
occurs, for example, in side-coupled double quantum dots
\cite{vojta2002, cornaglia2005, sidecoupled, hofstetter2002} and
triple quantum dot coupled in series \cite{tripike}. In parallelly
coupled systems, the residual interaction between the remaining spin
and the Fermi liquid quasi-particles is, however, ferromagnetic as can
be deduced from the splitting of the NRG energy levels in the
strong-coupling fixed point \cite{koller2005}: the strong-coupling
fixed point is stable.

We have thus demonstrated that with decreasing temperature the
symmetric ($\delta=0$) multi-impurity Anderson model flows from the FO
regime, through LM and FF regimes, to a stable underscreened $S=N/2$
Kondo model strong-coupling fixed point. The summary of different
regimes is given in Table~\ref{tab_regimes}.

\begin{table*}[htb]
\centering
\begin{ruledtabular}
\begin{tabular}{@{}cccccc@{}}
Regime             & Relevant states   & Magnetic susceptibility &
Spin correlations  & Charge fluctuations & Entropy \\
 &  & $\mu_\mathrm{eff}=T\chi_\mathrm{imp}(T)/(g\mu_B)^2$ &
$\langle \vc{S}_1 \cdot \vc{S}_2 \rangle$ &
$\langle n^2 \rangle-\langle n \rangle^2$ & $S_\mathrm{imp}$ \\
\colrule
FO & $N \times (\ket{0}, \ket{\uparrow}, \ket{\downarrow},
\ket{2})$ & $N/8$ & $0$ & $O(1)$ & $N\ln 4$ \\
LM & $N \times (\ket{\uparrow}, \ket{\downarrow})$ & $N/4$
& $0$ & small & $N \ln 2$ \\
FF & $\ket{S=N/2,S_z}$ & $N/2(N/2+1)/3$
& $\sim 1/4$ & small & $\ln(N+1)$ \\
SC & $\ket{S=N/2-1/2,S_z}$ & $(N-1)/2(N/2+1/2)/3$
& $\sim 1/4$ & small & $\ln N$ \\
\end{tabular}
\end{ruledtabular}
\caption{Regimes of the symmetric ($\delta=0$) $N$-impurity Anderson
model}
\label{tab_regimes}
\end{table*}

\section{Stability of $N=2$ systems with respect to various perturbations}
\label{sec_stability}

We next explore the effect of various physically relevant
perturbations with a special emphasis on the robustness of the
ferromagnetically frozen state and the ensuing $S=N/2$ Kondo effect
against perturbation of increasing strength. We show that the system
of multiple quantum dots remains in a $S=N/2$ state even for
relatively large perturbations. We also study the quantum phase
transitions from the $S=N/2$ state driven by strong perturbations. In
this sections we limit our calculations to the $N=2$ system.

\subsection{Variation of the on-site energy levels}

\subsubsection{Deviation from the particle-hole symmetric point}
\label{devidelta}

\begin{figure}[htbp!]
\includegraphics[width=8cm,clip]{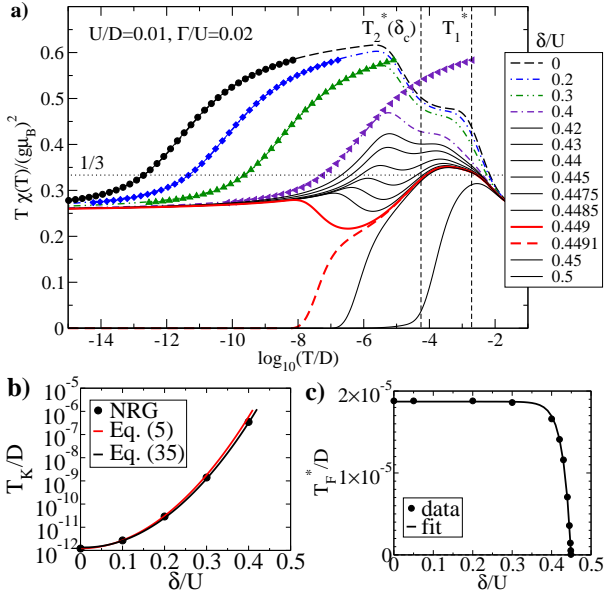}
\caption{(Color online) a) Temperature-dependent susceptibility of the
2-dot systems on departure ($\delta>0$) from the particle-hole
symmetric point, $\delta=0$.  Symbols are fits to the universal
susceptibility obtained using the Bethe-Ansatz method for the $S=1$ Kondo
model.  b) Calculated and predicted Kondo temperature,
Eq.~\eqref{TKJ_delta}. For comparison we also plot $T_K$ given by
Eq.~\eqref{tkh}, which shows expected discrepancy for large $\delta/U$.
c) Calculated $T_F^*$ and the fit to an exponential function. }

\label{fig_h}
\end{figure}

A small departure from the particle-hole symmetric point ($\delta \neq
0$) does not destabilize the $S=N/2$ Kondo behavior: the magnetic
susceptibility curves still follow the Bethe-Ansatz results even for
$\delta/U$ as large as $0.4$, see Fig.~\ref{fig_h}a. For
$\delta>\delta_c$, where $\delta_c/D \sim 0.45$ is the critical value
of parameter $\delta$, the triplet state is
destabilized. Consequently, there is no Kondo efect. This is a
particular case of the singlet-triplet transition that is a subject of
intense studies in recent years, both experimentally
\cite{entinwohlman2001, fuhrer2004a, kogan2003} and theoretically
\cite{izumida2001, pustilnik2001, pustilnik2003, hofstetter2002,
hofstetter2004}.

In the asymmetric single impurity model, the valence-fluctuation (VF) regime
is characterized by $\mu_\mathrm{eff} = T \chi(T) / (g\mu_B)^2 \sim 1/6$
\cite{krishna1980b}. The VF regimes occurs at $T_1^*$ and the transition
from VF to LM regime occurs at $T_2^* \sim |E_d^*|/\alpha$, where $E_d^*$ is
the renormalized on-site energy of the impurity: $E_d^* = \epsilon_d -
\frac{\Gamma}{\pi} \ln \left( -U/E_d^* \right)$. For two uncorrelated dots
in the VF regime, we expect $\mu_\mathrm{eff} \sim 1/3$. In
Fig.~\ref{fig_h}a we plotted a number of susceptibility curves for
parameters $\delta$ in the proximity of the singlet-triplet transition.
While there is no clearly-observable valence-fluctuation plateau, the value
of $\mu_\mathrm{eff}$ is indeed near $1/3$ between $T_1^*$ and
$T_2^*(\delta_c)$.

In Fig.~\ref{fig_h}b we compare calculated Kondo temperatures with
analytical predictions based on the results for the single impurity model
\cite{krishna1980b}. For $\delta \neq 0$, $J_\mathrm{K}$ generalizes
according to the Schrieffer-Wolff transformation, Eq.~\eqref{JKdelta}.
Departure from the p-h symmetric point also induces potential scattering
\begin{equation}
\rho_0 K = \frac{\Gamma}{2\pi} \left(
\frac{1}{|\delta - U/2|} - \frac{1}{|\delta + U/2|} 
\right).
\end{equation}
The effective ${\tilde J}_\mathrm{K}$ that enters the expression for the
Kondo temperature is \cite{krishna1980b}
\begin{equation}
{\tilde J}_\mathrm{K}=J_\mathrm{K} \left[
1+ \left( \pi \rho_0 K \right)^2 \right],
\end{equation}
and the effective bandwidth $0.182 U$ is replaced by $0.182 |E_d^*|$. The
Kondo temperature is now given by
\begin{equation}
\label{TKJ_delta}
T_K = 0.182 |E_d^*| \sqrt{\rho_0 {\tilde J}_\mathrm{K}} 
\exp \left( -1 / (\rho_0 {\tilde J}_\mathrm{K})\right).
\end{equation}
This analytical estimate agrees perfectly with the NRG results: for
moderate $\delta/U$, the results obtained for asymmetric single
impurity model also apply to the multi-impurity Anderson model.

In Fig.~\ref{fig_h}c we show the $\delta$-dependence of the LM-FF transition
temperature $T_F^*$. Its value remains nearly independent of $\delta$ in the
interval $\delta \lesssim 0.4 U$ and then it suddenly drops. More
quantitatively, the dependence on $\delta$ can be adequately described using
an exponential function
\begin{equation}
T_F^*(\delta) = T_F^*(0)
\left[ 1-\exp\left( \frac{\delta-\delta_c}{\lambda} \right) \right]
\end{equation}
where $T_F^*(0)/D=1.8 \times 10^{-5}$ is the transition temperature in
the symmetric case, $\delta_c/D=0.45$ is the critical $\delta$ and
$\lambda/D=2.1 \times 10^{-2}$ is the width of the transition region.
Exchange interaction $J_\mathrm{RKKY}$ does not depend on $\delta$ for
$U/D=0.01 \ll 1$, which explains constant value of $T_F^*(\delta)$ for
$\delta \lesssim 0.4 U$. At a critical value $\delta_c$, $T_F^*$ goes
to zero and for still higher $\delta$ the spin-spin correlation
becomes antiferromagnetic. Since the ground-state spins are different,
the triplet and singlet regime are separated by a quantum phase
transition at $\delta=\delta_c$. This transition is induced by charge
fluctuations which destroy the ferromagnetic order of spins as the
system enters the VF regime.  The exponential dependence arises from
the grand-canonical statistical weight factor $\exp[\delta (n-2)/(k_B
T)]$, where $n$ is the number of the electrons confined on the dots.
The transition is of the first order, since for equal coupling of both
impurities to the band there is no mixing between the $n=2$ triplet
states and the $n=0$ singlet state \cite{vojta2002}.

For $\delta$ slightly lower than the critical $\delta_c$, the
effective moment $T\chi(T)$ shows a rather unusual temperature
dependence. It first starts decreasing due to charge fluctuations,
however with further lowering of the temperature the moment ordering
wins over, $T \chi(T)$ increases and at low-temperatures approaches
the value characteristic for the partially screened $S=1$ moment,
i.e. $T\chi/(g\mu_B)^2 \sim 1/4$.

\subsubsection{Splitting of the on-site energy levels}

\begin{figure}[htbp!]
\includegraphics[width=8cm,clip]{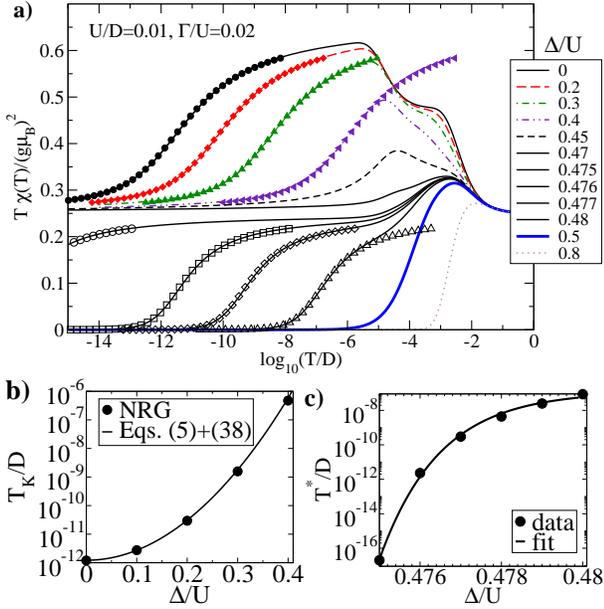}
\caption{(Color online) a) Temperature-dependent susceptibility of 
the $2$-dot system with unequal (detuned) on-site energies, 
$\delta_1 = \Delta$, $\delta_2 = - \Delta$. 
Full symbols present Bethe-Ansatz results of the equivalent $S=1$ Kondo model,
while empty symbols are BA results of a $S=1/2$ Kondo model.
b) Comparison of calculated and
predicted Kondo temperature, see Eqs.~\eqref{tkh} and ~\eqref{jkDelta}.
c) The Kondo temperature of the S=1/2 Kondo screening on the
singlet side of the transition and a fit to Eq.~\eqref{tk2delta}.
}
\label{fig_k}
\end{figure}

We next consider the 2-dot Hamiltonian with unequal on-site energies
$\delta_i$:
\begin{equation}
H_{\mathrm{dot},i} = \delta_i (n_i-1) + \frac{U}{2}(n_i-1)^2.
\end{equation}
We focus on the case $\delta_1 = \Delta$ and $\delta_2 = - \Delta$,
which represents another experimentally relevant perturbation. This
model is namely particle-hole symmetric for an arbitrary choice of
$\Delta$ under a generalized p-h transformation $c^\dag_{k\sigma} \to
c_{k,-\sigma}$, $d^\dag_{1\sigma} \to d_{2,-\sigma}$,
$d^\dag_{2\sigma} \to d_{1,-\sigma}$. The total occupancy of both dots
is exactly 2 for any $\Delta$. We can therefore study the effect of
the on-site energy splitting while maintaining the particle-hole
symmetry. Susceptibility curves are shown in Fig.~\ref{fig_k}a for a
range of values of $\Delta$. For $\Delta$ up to some critical value
$\Delta_c\approx 0.47$ the 2-dot Anderson model remains equivalent to
the $S=1$ Kondo model for $T<T_F^*$. A singlet-triplet transition of
the Kosterlitz-Thouless type \cite{hofstetter2002, vojta2002} occurs
at $\Delta=\Delta_c$.

Even though the two dots are now inequivalent, the Schrieffer-Wolff
transformation yields the same $J_K$ for both spin impurities. We obtain
\begin{equation}
\label{jkDelta}
J_K = 2 |V_{k_F}|^2 
\left( \frac{1}{|\Delta-U/2|}+\frac{1}{|\Delta+U/2|} \right).
\end{equation}
Due to the particle-hole symmetry no potential scattering is generated. The
effective Kondo Hamiltonian for small $\Delta$ is thus nearly the same as 
that for small $\delta$ discussed in the previous section. In
Fig.~\ref{fig_k}b calculated Kondo temperatures are plotted in comparison
with analytical result from Eqs.~\eqref{tkh} and \eqref{jkDelta}. The
agreement is excellent.

The properties of the systems with $\delta \neq 0$ and $\Delta \neq 0$
become markedly different near respective singlet-triplet transition
points. For $\delta \neq 0$, the transition is induced by charge
fluctuations which suppress magnetic ordering and, due to equal
coupling of both dots to the band, the transition is of first
order. For $\Delta \neq 0$ the transition is induced by depopulating
dot 2 and populating dot 1 while the total charge on the dots is
maintained, which leads to the transition from an inter-impurity
triplet to a local spin-singlet on the dot 1. Since there is an
asymmetry between the dots, the transition is of the
Kosterlitz-Thouless type \cite{vojta2002}.

The Kondo temperature of the $S=1/2$ Kondo screening near the
transition on the singlet side, $T^*$, is approximately given by
\begin{equation}
\label{tk2delta}
\log{T^*/D} = - \alpha - \beta \exp\left(-\frac{\Delta-{\tilde \Delta}}
{\lambda} \right).
\end{equation}
We obtain $\alpha \approx 7$, $\beta \approx 2.8$, ${\tilde \Delta}/D
\approx 0.477$ and $\lambda/D \approx 1.5 \times 10^{-3}$. This
expression is consistent with the cross-over scale formula $T^*
\propto \exp[-T_K/J_{12}]$ for a system of two fictitious spins, one
directly coupled to the conduction band and the other side-coupled to
the first one with exchange-interaction $J_{12}$ that depends
exponentially on $\Delta$: $J_{12} = T_K/\beta \exp[
(\Delta-{\tilde\Delta}) / \lambda]$.

\subsection{Inter-impurity interaction}
\subsubsection{Inter-impurity exchange interaction} \label{IVc}

In this subsection we show that by introducing an explicit exchange
interaction $J_{12}$ between the localized spins on the dots, the strength
of the RKKY interaction, $J_{\mathrm{RKKY}}$, can be directly determined. We
thus study the two-impurity Anderson model with
\begin{equation*}
H_{\mathrm{dots}} = \sum_{i=1}^2 H_{\mathrm{dot},i} + J_{12} \vc{S}_1 \cdot
\vc{S}_2,
\end{equation*}
where $J_{12}>0$.

\begin{figure}[htbp!]
\includegraphics[width=8cm,clip]{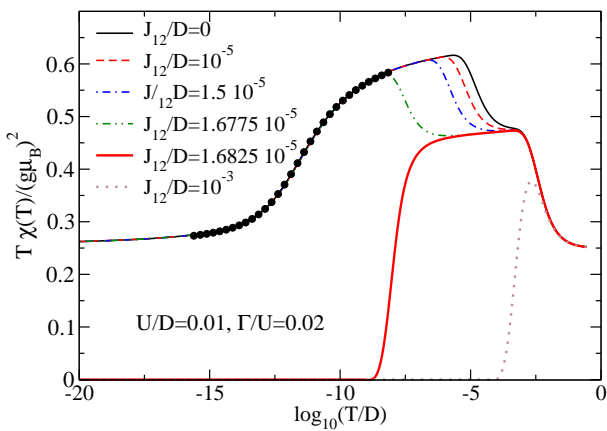}
\caption{(Color online) Temperature-dependent susceptibility of
the 2-dot systems for different anti-ferromagnetic inter-impurity
couplings $J_{12}$. Circles are Bethe-Ansatz results for the
susceptibility of the $S=1$ Kondo model with the Kondo temperature
which is equal for all cases where $J_{12} < J_\mathrm{c}$.}
\label{fig_c}
\end{figure}

As seen from Fig.~\ref{fig_c}, for $J_{12}$ above a critical value
$J_c$, the RKKY interaction is compensated, local moments on the dots
form the singlet rather than the triplet which in turn prevents
formation of the $S=1$ Kondo effect. The phase transition is of the
first order
\cite{vojta2002}. Using Eq.~\eqref{extract}, we obtain $T_F^*/D
\approx 1.87 \times 10^{-5}$ for the non-perturbed problem with the
same $U$ and $\Gamma$, while $J_c/D \approx 1.68 \times
10^{-5}$. Taking into account the definition $T_F^* =
J_\mathrm{RKKY}/\beta$, where $\beta \sim 1$, we conclude that
$J_\mathrm{RKKY}$ agrees well with the critical value of $J_c$, {\it
i.e.} $J_c = J_\mathrm{RKKY}$.
The perturbation theory prediction of $J_\mathrm{RKKY}/D = 1.6 \times
10^{-5}$ also agrees favorably with numerical results.

As long as $J_{12} < J_c$, even for $J_{12}>T_K$, the $S=1$ Kondo effect
survives and, moreover, the Kondo temperature remains unchanged, determined
only by the value of $\rho_0 J_\mathrm{K}$ as in the $J_{12}=0$ case. The
only effect of increasing $J_{12}$ in the regime where $J_{12}<J_\mathrm{c}$
is the reduction of the transition temperature into the triplet state, which
is now given by $T_F^* \sim J_\mathrm{eff} / \beta$ with the effective
inter-impurity interaction $J_\mathrm{eff} = J_\mathrm{RKKY}-J_{12}$.

\subsubsection{Hopping between the impurities}

\begin{figure}[htbp!]
\includegraphics[width=8cm,clip]{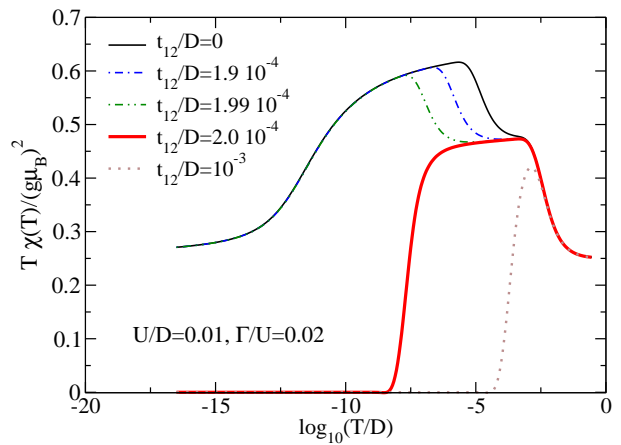}
\caption{(Color online) Temperature-dependent susceptibility of the 
2-dot systems with inter-dot tunneling coupling $t_{12}$.
For $t_{12}/D \approx 2 \times 10^{-4}$, we have
$J_\mathrm{AFM}/D \approx 1.6 \times 10^{-5}$, which
agrees well with the critical value of $J_c/D \approx 1.7 \times
10^{-5}$ found in the case of an explicit exchange interaction
between the dots, see Fig.~\ref{fig_c}.
}
\label{fig_j}
\end{figure}

We now study the two-impurity Anderson model with additional
hopping between the dots:
\begin{equation}
H_{\mathrm{dots}} = \sum_{i=1}^2 H_{dot,i} -
t_{12} \sum_\sigma \left( d^\dag_{1\sigma} d_{2\sigma} + d^\dag_{2\sigma}
d_{1\sigma} \right),
\end{equation}
This model can be viewed also as a  single-channel version of the
Alexander-Anderson model \cite{alexander1964} in the limit of zero
separation between the impurities. The magnetic-susceptibility
curves are shown in Fig.~\ref{fig_j}.

The hopping leads to hybridization between the atomic levels of the dots
which in turn results in the formation of an even and odd level (``molecular
orbital'') with energies $\epsilon_{e,o}= \epsilon_d \pm t_{12}$. In the
presence of interaction $U$ there are two contributions to the energy of the
low-lying states: ``orbital energy'' proportional to $t_{12}$ and ``magnetic
energy'' due to an effective antiferromagnetic exchange
$J_{\mathrm{AFM}}=4t_{\mathrm{12}}^2/U$, which is second-order in $t_{12}$.
Even though the orbital energy is the larger energy scale, the Kondo
effect is largely insensitive to the resulting level splitting. Instead, the
Kondo effect is destroyed when $J_\mathrm{AFM}$ exceeds $J_\mathrm{RKKY}$,
much like in the case of explicit exchange interaction between the dots
which was discussed in the previous subsection. We should emphasize the
similarity between the curves in Figs.~\ref{fig_c} and \ref{fig_j}.

In the wide-band limit $U \ll D$, $J_\mathrm{RKKY}/D \approx 0.62 \times 16
V^4/U$, therefore the critical $t_{12,c}$ is given by $t_{12,c} \approx
\Gamma$ and it does not depend on $U$. This provides an alternative
interpretation for the $U$-dependence of $t_{12,c}$ in the strong-coupling
regime found in Ref.~\onlinecite{nishimoto2006}.

\subsection{Isospin-invariance breaking perturbations}

The inter-impurity electron repulsion and the two-electron hopping between
the impurities represent perturbations that break the isospin $SU(2)$
symmetry of the original model, while they preserve both the particle-hole
symmetry as well as the spin invariance.

\subsubsection{Inter-impurity electron repulsion}

The effect of the inter-impurity electron repulsion (induced by
capacitive coupling between the two parallel quantum dots) is studied
using the Hamiltonian
\begin{equation}
H_{\mathrm{dots}} = \sum_{i=1}^2 H_{dot,i} + U_{12} (n_1-1) (n_2-1),
\end{equation}
where it should be noted that $(n_1-1)(n_2-1)=4 I_1^z I_2^z$ is the
longitudinal part of the isospin-isospin exchange interaction $\vc{I}_1
\cdot \vc{I}_2$.

\begin{figure}[htbp!]
\includegraphics[width=8cm,clip]{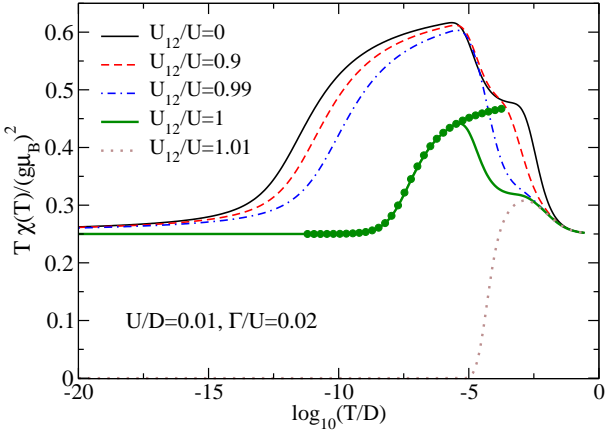}
\caption{(Color online) Temperature-dependent susceptibility of
the 2-dot systems for different inter-impurity electron-electron
repulsion parameters $U_{12}$. Circles are the Bethe-Ansatz results 
for the $S=1/2$ Kondo model which fit the NRG results in the special
case $U_{12}=U$.} \label{fig_d}
\end{figure}

Results in Fig.~\ref{fig_d} show that the inter-impurity repulsion is not an
important perturbation as long as $U_{12} < U$. Finite $U_{12}$ only
modifies the Kondo temperature and the temperature $T_1^*$ of the FO-LM
transition, while the behavior of the system remains qualitatively
unchanged. Note that $T_F^*$ is unchanged since $U_{12}$ equally affects
both the singlet and the triplet energy.

For $U_{12}>U$ the electrons can lower their energy by forming on-site
singlets and the system enters the {\it charge-ordering regime}
\cite{galpin2005}. This behavior bares some resemblance to that of the
negative-$U$ Anderson model \cite{taraphder1991} which undergoes a charge
Kondo effect.

\begin{figure}[htbp!]
\includegraphics[width=8cm,clip]{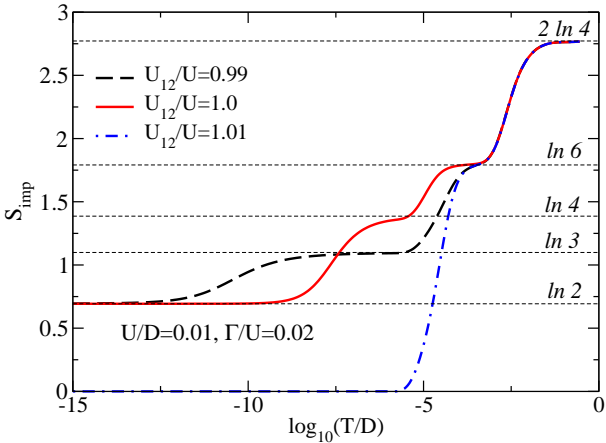}
\caption{(Color online) Temperature-dependent entropy of the 2-dot 
systems for different inter-impurity electron-electron repulsion $U_{12}$.} 
\label{fig_ds}
\end{figure}

The system behaves in a peculiar way at the transition point $U_{12}=U$
where $U_{12}$ and $U$ terms can be combined using isospin operators as
\begin{equation}
\label{U12sim}
U/2 \left( 4 (I_1^z)^2 + 4 (I_2^z)^2 \right) + U_{12} 4 I_1^z I_2^z
= 2U (I^z)^2.
\end{equation}
We now have an intermediate temperature fixed point with a six-fold symmetry
of states with $I_z=0$ as can be deduced from Eq.~\eqref{U12sim} and the
entropy curve in Fig.~\ref{fig_ds}. 

For two impurities we can define an orbital pseudo-spin operator as
\begin{equation}
\vc{O} = \frac{1}{2} \sum_\alpha
\sum_{i,j=1,2}d^\dagger_{i\alpha} \boldsymbol{\sigma}_{ij} d_{j\alpha},
\label{izospin}
\end{equation}
where $\boldsymbol{\sigma}$ is the vector of the Pauli matrices. The quantum
dots Hamiltonian $H_{\mathrm{dots}}$ commutes for $U_{12}=U$ with all three
components of the orbital pseudo-spin operator; the decoupled
impurities thus have orbital $SU(2)_\mathrm{orb}$ symmetry. Furthermore,
pseudo-spin $\vc{O}$ and spin $\vc{S}$ operators commute and the symmetry is
larger, $SU(2)_\mathrm{spin} \otimes SU(2)_\mathrm{orb}$. In fact, the set
of three $S^i$, three $O^i$ and nine operators $S^i O^j$ are the generators
of the $SU(4)$ symmetry group of which $SU(2)_\mathrm{spin} \otimes
SU(2)_\mathrm{orb}$ is a subgroup. The six degenerate states are the spin
triplet, orbital singlet and the spin singlet, orbital triplet
\cite{deleo2004} which form a $SU(4)$ sextet:
\begin{equation*}
\label{sextuplet}
\begin{split}
\ket{S=1,S_z=1,O=0} &= \ket{\uparrow,\uparrow}, \\
\ket{S=1,S_z=0,O=0} &= 1/\sqrt{2} \left( \ket{\uparrow,\downarrow} +
\ket{\downarrow,\uparrow} \right), \\
\ket{S=1,S_z=-1,O=0} &= \ket{\downarrow,\downarrow}, \\
\ket{S=0,O=1,O_z=1} &= \ket{\uparrow\downarrow,0}, \\
\ket{S=0,O=1,O_z=0} &= 1/\sqrt{2} \left( \ket{\uparrow,\downarrow}
-
\ket{\downarrow,\uparrow} \right), \\
\ket{S=0,O=1,O_z=-1} &= \ket{0,\uparrow\downarrow}.
\end{split}
\end{equation*}
The states $\ket{S=0,O=1,O_z=\pm 1}$ can be combined into an isospin triplet
$\ket{S=0,I=1,I_z=0} = 1/\sqrt{2} \left(\ket{\uparrow\downarrow,0} +
\ket{0,\uparrow\downarrow} \right)$ and an isospin singlet $\ket{S=0,I=0} =
1/\sqrt{2} \left( \ket{\uparrow\downarrow,0} - \ket{0,\uparrow\downarrow}
\right)$.

The coupling of impurities to the leads, however, breaks the orbital
symmetry. Unlike the model studied in Ref.~\onlinecite{galpin2005}, our
total Hamiltonian $H$ is not $SU(4)$ symmetric, so no $SU(4)$ Kondo effect
is expected. Instead, as the temperature decreases the degeneracy first
drops from 6 to 4 and then from 4 to 2 in a $S=1/2$ $SU(2)$ Kondo effect
(see the fit to the Bethe-Ansatz result in Fig.~\ref{fig_d}). There is a
residual two-fold degeneracy in the ground state. To understand these
results, we applied perturbation theory (Appendix~\ref{app_rkky}) which
shows that the sextuplet splits in the fourth order perturbation in $V_k$.
The spin-triplet states and the state $\ket{S=0,I=0}$ form the new four-fold
degenerate low-energy subset of states, while the states
$\ket{S=0,I=1,I_z=0}$ and $\ket{S=0,O=1,O_z=0}$ have higher energy. The
remaining four states can be expressed in terms of even and odd
molecular-orbitals described by operators $d_{e\sigma}^\dag = 1/\sqrt{2}
\left( d^\dag_{1\sigma} + d^\dag_{2\sigma} \right)$ and $d_{o\sigma}^\dag =
1/\sqrt{2} \left( d^\dag_{1\sigma} - d^\dag_{2\sigma} \right)$. We obtain
\begin{equation}
\begin{split}
\ket{S=1,S_z=1,O=0} &= d^\dag_{e,\uparrow} d^\dag_{o,\uparrow} \ket{0}, \\
\ket{S=1,S_z=0,O=0} &= 1/\sqrt{2} \left( d^\dag_{o,\uparrow}
d^\dag_{e,\downarrow} + d^\dag_{e,\uparrow} d^\dag_{o,\downarrow} \right)
\ket{0}, \\
\ket{S=1,S_z=-1,O=0} &= d^\dag_{e,\downarrow} d^\dag_{o,\downarrow} \ket{0},\\
\ket{S=0,I=0} &= 1/\sqrt{2} \left( d^\dag_{o,\uparrow}
d^\dag_{e,\downarrow} - d^\dag_{e,\uparrow} d^\dag_{o,\downarrow} \right)
\ket{0}.
\end{split}
\end{equation}
The four remaining states are therefore a product of a spin-doublet in the
even orbital and a spin-doublet in the odd orbital. Due to the symmetry of
our problem, only the even orbital couples to the leads, while the odd
orbital is entirely decoupled. The electron in the even orbital undergoes
$S=1/2$ Kondo screening, while the unscreened electron in the odd orbital is
responsible for the residual two-fold degeneracy.

\subsubsection{Two-electron hopping}

We consider the Hamiltonian
\begin{equation}
H_{\mathrm{dots}} = \sum_{i=1}^2 H_{dot,i} - T_{12} \hat{T},
\end{equation}
where $\hat{T}$ is the two-electron hopping operator that can be
expressed in terms of  the transverse part of the isospin-isospin
exchange interaction $\vc{I}_1 \cdot \vc{I}_2$:
\begin{equation}
\begin{split}\label{pairhop}
\hat T &= d^\dag_{1\uparrow} d^\dag_{1\downarrow}
d_{2\downarrow} d_{2\uparrow} + 
d^\dag_{2\uparrow} d^\dag_{2\downarrow}
d_{1\downarrow} d_{1\uparrow}
\\
&= I_1^+ I_2^- + I_1^- I_2^+ = 2(I_1^x I_2^x + I_1^y I_2^y).
\end{split}
\end{equation}
This perturbation term is complementary to the one generated by $U_{12}$ in
Eq.~\eqref{U12sim} and studied in the previous subsection. Physically, it
corresponds to correlated tunneling of electron pairs which can be neglected
in the applications to problems of transport through parallel quantum dots
coupled electrostatically as physically realized in semiconductor
heterostructures. Models featuring pair-tunneling terms as in
Eq.~\eqref{pairhop} may, however, be of interest to problems in tunneling
through molecules with vibrational degrees of freedom, where ground states
with even number of electrons can be favored due to a polaronic energy shift
\cite{mravlje2005, koch2006}. In such cases, the charge transport is
expected to be dominated by the electron-pair tunneling \cite{koch2006}.

\begin{figure}[htbp!]
\includegraphics[width=8cm,clip]{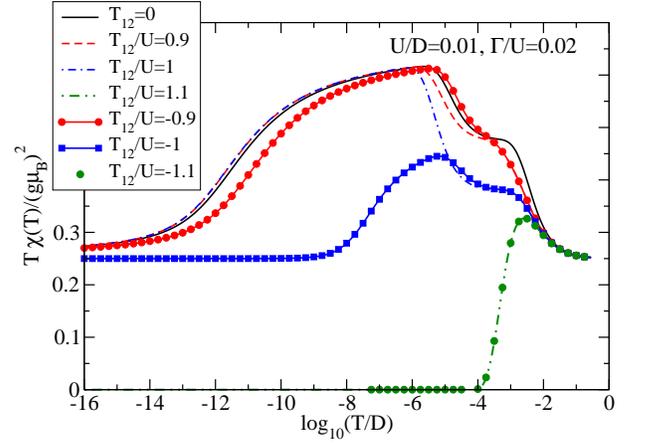}
\caption{(Color online) Temperature-dependent susceptibility of the
$2$-dot system with two-electron hopping between the dots
$T_{12}$.} \label{fig_m}
\end{figure}

The temperature dependence of the magnetic susceptibility shown in
Fig.~\ref{fig_m} again demonstrates the robustness of the $S=1$ state for
$|T_{12}| < U$. The behavior of the system for negative $T_{12}$ is similar
to the case of the inter-impurity repulsion. For $T_{12}=-U$ we again
observe special behavior of the susceptibility curve, characteristic for the
six-fold degeneracy observed in the previous subsection at $U_{12}=U$. For
positive $T_{12}$ the system undergoes the $S=1$ spin Kondo effect up to and
{\it including} $T_{12}=U$. The FO-LM transition temperature $T_1^*$ and the
Kondo temperature are largely $T_{12}$ independent, while the LM-FF
transition temperature $T_F^*$ decreases with increasing $T_{12}$.

\subsection{Unequal coupling to the continuum}
\label{unequal}

We finally study the Hamiltonian that allows for unequal
hybridizations $\Gamma_i = \pi \rho_0 |V^i_{k_F}|^2$ in the following
form:
\begin{equation}
H = H_\mathrm{band} + H_{\mathrm{dots}} + \sum_{i=1}^2 H_{\mathrm{c},i},
\end{equation}
with
\begin{equation}
H_{\mathrm{c},i}=\frac{1}{\sqrt{L}} \sum_{k\sigma} \left(
V_k^i d^\dag_{i\sigma} c_{k\sigma} + \text{H.c.} \right).
\end{equation}
We set $V^2_k=\alpha V^1_k$, i.e. $\Gamma_2=\alpha^2 \Gamma_1$.

The effective low-temperature Hamiltonian can be now written as
\begin{equation}\label{heffalf}
H_\mathrm{eff} = H_\mathrm{band} + \vc{s} \cdot \sum_{i=1}^2
J_{\mathrm{K},i} \vc{S}_i - J^{\mathrm{eff}}_{\mathrm{RKKY}} \vc{S}_1 \cdot
\vc{S}_2.
\end{equation}
with $J_{K,2}=\alpha^2 J_{K,1}$ and with the effective RKKY
exchange interaction given by a generalisation of Eq.~\eqref{rkkydep}
\begin{equation}\label{RKKYeff}
J^{\mathrm{eff}}_{\mathrm{RKKY}} = 0.62 U \rho_0^2
J_{\mathrm{K},1} J_{\mathrm{K,2}} = \alpha^2 J_{\mathrm{RKKY}},
\end{equation}
where $J_{\mathrm{RKKY}}$ is the value of RKKY parameter at
$\alpha=1$.  In our attempt to derive the effective Hamiltonian we
assume that in the temperature regime $T \lesssim
J^{\mathrm{eff}}_\mathrm{RKKY}$ the two moments couple into a
triplet. Since the two Kondo exchange constants $J_{\mathrm{K},i}$ are
now different, we rewrite $H_\mathrm{eff}$ in Eq.~\eqref{heffalf} in
the following form
\begin{equation}
\begin{split}
H_\mathrm{eff} = H_\mathrm{band} &+ \vc{s}
\cdot \left( \frac{J_{\mathrm{K},1}+J_{\mathrm{K},2}}{2} \left(
\vc{S}_1+\vc{S}_2 \right) \right)\\
&+ \vc{s}
\cdot \left( \frac{J_{\mathrm{K},1}-J_{\mathrm{K},2}}{2} \left(
\vc{S}_1-\vc{S}_2 \right) \right) \\
&- J^{\mathrm{eff}}_{\mathrm{RKKY}}\vc{S}_1 \cdot \vc{S}_2.
\end{split}
\end{equation}

Within the triplet subspace, $\vc{S}_1+\vc{S}_2$ is equal to the new
composite spin 1, which we denote by $\vc{S}$, $\vc{S}_1-\vc{S}_2$ is
identically equal to zero, and $\vc{S}_1 \cdot \vc{S}_2$ is a constant
$-1/4$. As a result, the effective $J_{\mathrm{K}}$ is simply the
average of the two exchange constants:
\begin{equation}
\label{Javg}
J_{\mathrm{K},\mathrm{eff}}
= \frac{ J_{\mathrm{K},1} + J_{\mathrm{K},2}}{ 2 }.
\end{equation}

\begin{figure}[htbp!]
\includegraphics[width=8cm,clip]{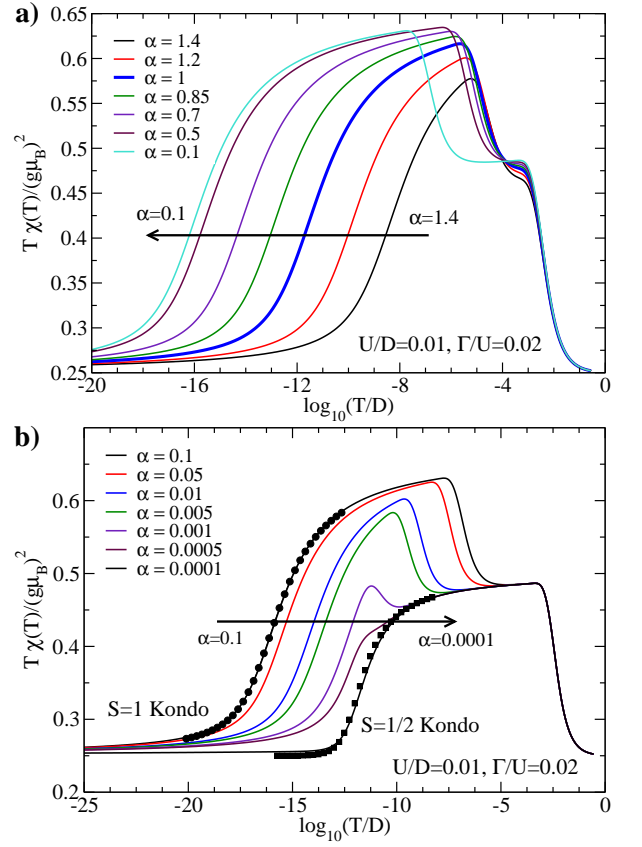}
\caption{(Color online) Temperature-dependent susceptibility of the
$2$-dot system with unequal coupling to the leads, $\Gamma_2 =
\alpha^2 \Gamma_1$. a) The range of $\alpha$ where $T_K$ is decreasing. 
b) The range of $\alpha$ where $T_K$ is increasing again. Circles
(squares) are BA results for the $S=1$ ($S=1/2$) Kondo model.
The arrows indicate the evolution of the susceptibility curves as
the parameter $\alpha$ decreases.
} 
\label{fig_l}
\end{figure}

\begin{figure}[htbp!]
\includegraphics[width=8cm,clip]{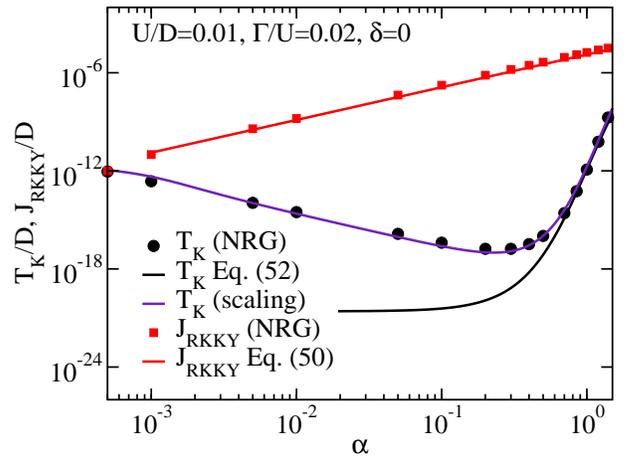}
\caption{(Color online) Comparison of calculated and predicted 
Kondo temperature $T_K$ and effective exchange interaction 
$J^\mathrm{eff}_\mathrm{RKKY}$.
The calculation of scaling results for $T_K$ is described in
Appendix~\ref{scaling}.
}
\label{fig_r4}
\end{figure}

Susceptibility curves for different $\alpha$ are shown in
Fig.~\ref{fig_l}.  Note that the Kondo temperature determined using
Eq.~\eqref{tkh} combined with the naive argument given in
Eq.~\eqref{Javg} fails to describe the actual Kondo scale for $\alpha
\lesssim 0.4$ as seen from Fig.~\ref{fig_r4}.  This is due to
admixture of the singlet state, which also renormalizes $J_{K}$, even
though the singlet is separated by $J^{\mathrm{eff}}_{\mathrm{RKKY}}
\gg T_K$ from the triplet subspace. Note however, that
$J^{\mathrm{eff}}_{\mathrm{RKKY}}$ is well described by the simple
expression given in Eq.~\eqref{RKKYeff} as shown in
Fig.~\ref{fig_r4}. By performing a second-order RG calculation (see
Appendix~\ref{scaling}), which takes the admixture of the singlet state
into account, we obtain $T_K$ as a function of $\alpha$ which agrees
very well with the NRG results, see Fig.~\ref{fig_r4}.

For extremely small $\alpha$, $J^{\mathrm{eff}}_{\mathrm{RKKY}}$
eventually becomes comparable to the Kondo temperature, see
Fig.~\ref{fig_r4}. For that reason the ferromagnetic locking-in is
destroyed and the system behaves as a double $S=1/2$ doublet, one
of which is screened at $T_K^1=T_K(J_{\mathrm{K,1}})$ as shown in
Fig.~\ref{fig_l}b.

\section{Conclusions}

We have shown that several magnetic impurities, coupled to the same
Wannier orbital of a conduction electron band, experience
ferromagnetic RKKY interaction which locks local moments in a state of
a maximal total spin. The multi-impurity Anderson model is at low
temperatures, {\it i.e.} for $T<T_F^*$, equivalent to a $S=N/2$
$SU(2)$ Kondo model. Using perturbation theory up to the fourth order
in $V$ we derived an analytical expression for $J_{\mathrm{RKKY}}$ and
tested it against NRG calculations.  We have also shown that the
high-spin state is very robust against experimentally relevant
perturbations such as particle-hole symmetry breaking, on-site energy
level splitting, inter-impurity capacitive coupling and direct
exchange interaction. At low temperatures, the ferromagnetically
locked impurities undergo a collective Kondo cross-over in which half
of a unit of spin is screened. The Kondo temperature in this simple
model does not depend on the total spin (i.e. on the number of
impurities $N$), while the LM-FF temperature $T_F^*$ is weakly
$N$-dependent.

We next list a few most important findings concerning the effect of
various perturbations to the original two-dot system: a)
$T_F^*$ is in the range $\delta \lesssim 0.4 U$ nearly independent of
the deviation from the particle-hole symmetric point $\delta=0$,
b) increasing the difference between on-site energies of two dots,
$2\Delta$, induces a Kosterlitz-Thouless type phase transition
separating the phase with $S=1/2$ residual spin at low-temperatures
from the $S=0$ one, c) introduction of additional one-electron hopping
between the impurities induces effective AFM interaction
$J_{\mathrm{AFM}}=4t_{12}^2/U$ that does not effect the Kondo
temperature as long as $J_{\mathrm{AFM}}\lesssim J_{\mathrm{RKKY}}$,
nevertheless, at $t_{12}=t_{12,c}$ it destabilizes the $S=1$
state. The critical value $t_{12,c}\sim \Gamma$ does not depend on
$U$, d) inter-impurity Coulomb interaction $U_{12}$ leads to a
transition from the $S=1$ Kondo state to the charge ordered state. In
the 4-fold degenerate intermediate point, reached at $U_{12}=U$, the
effective Hamiltonian consists of the effective $S=1/2$ Kondo model
and of a free, decoupled $S=1/2$ spin, e) when the two impurities are
coupled to the leads with different hybridization strengths,
second-order scaling equations provide a good description of the Kondo
temperature.

The properties of our model apply very generally, since high-spin
states can arise whenever the RKKY interaction is ferromagnetic, even
when the dots are separated in space \cite{usaj2005, tamura2005}. In
addition, it has become possible to study Kondo physics in clusters of
magnetic atoms on metallic surfaces \cite{jamneala2001,
aligia2006}. On (111) facets of noble metals such as copper, bulk
electrons coexist with Shockley surface-state electrons
\cite{lin2005}. Surface-state bands on these surfaces have $k_F \sim
0.1 - 0.2 \AA^{-1}$; thus, for nearest and next-nearest neighbor
adatoms $k_F R \lesssim 1$. If hybridization to the surface band is
dominant, small clusters then effectively couple to the same Wannier
orbital of the surface band and the single-channel multi-impurity
Anderson model is applicable; in the absence of additional
inter-impurity interactions, the spins would then tend to order
ferromagnetically.  If hybridization to the bulk band with
$k^{3\mathrm{D}}_F \sim 1 \AA^{-1}$ is also important, the problem
must be described using a complex two-band multi-channel Hamiltonian.

Further aspects of the multi-impurity Anderson model should be
addressed in the future work. Systems of coupled quantum dots and
magnetic impurities on surfaces are mainly characterized by measuring
their transport properties. Conductance can be determined by
calculating the spectral density functions using the numerical
renormalization group method. We anticipate that the fully screened
$N=1$ model will have different temperature dependence as the
under-screened $N \geq 2$ models. Since in quantum dots the impurity
level $\delta$ (or $\epsilon_d$) can be controlled using gate
voltages, it should be interesting to extend the study to asymmetric
multi-impurity models for $N > 2$ where more quantum phase transitions
are expected in addition to the one already identified for $N=2$ at
$\delta=\delta_c$.

\begin{acknowledgments}
The authors acknowledge useful discussions with J. Mravlje
and L. G. Dias da Silva and the financial support of the SRA
under Grant No. P1-0044.
\end{acknowledgments}

\appendix

\section{Rayleigh-Schroedinger perturbation theory in $V_k$}
\label{app_rkky}

Following Ref.~\onlinecite{fye1990}, we apply Rayleigh-Schr\"odinger
perturbation theory to calculate second and fourth order corrections in
$V_k$ to the energy of a state $\ket{n}$:
\begin{equation}
\begin{split}
E_n^{(2)} &= {\sum_{m}}' \frac{\bra{n}H_c\ket{m} \bra{m}H_c\ket{n}}{E_n - E_m} \\
E_n^{(4)} &= {\sum_{m_1,m_2,m_3}}'  \\
&\frac{\bra{n}H_c\ket{m_3} \bra{m_3}H_c\ket{m_2} \bra{m_2}H_c\ket{m_1}
\bra{m_1}H_c\ket{n}} {(E_n-E_{m_3})(E_n-E_{m_2})(E_n-E_{m_1})} \\
\end{split}
\end{equation}
The summation extends over all intermediate states $\ket{m_i}$ not
equal to one of the degenerate ground states. We will consider the
simplified case of constant $V_k$, i.e. $V_k=V_{k_F}=V$.

\subsection{RKKY interaction in the two-impurity case}

We study the splitting between the singlet $\ket{S}=1/\sqrt{2} \left(
\ket{\uparrow,\downarrow}-\ket{\downarrow,\uparrow}\right)$ and the triplet
state $\ket{T}=1/\sqrt{2} \left( \ket{\uparrow,\downarrow} +
\ket{\downarrow,\uparrow}\right)$. The second order corrections are
$E_S^{(2)} = E_T^{(2)} = - (S_1 + S_2)$ with
\begin{equation}
\begin{split}
S_1 &= 4 V^2 \frac{1}{L} \sum_{k' \geq k_F} \frac{1}{U-2\delta+2 \epsilon_{k'}}, \\
S_2 &= 4 V^2 \frac{1}{L} \sum_{k \leq k_F} \frac{1}{U+2\delta-2 \epsilon_k}.
\end{split}
\end{equation}
There is therefore no splitting to this order in $V$. The fourth order
corrections are
\begin{equation}
\begin{split}
E_S^{(4)} &= W_S^{\mathrm{ph}} + W_S^{\mathrm{pp}} + W_S^{\mathrm{hh}}, \\
E_T^{(4)} &= W_T^{\mathrm{ph}},
\end{split}
\end{equation}
where the particle-hole ($\mathrm{ph}$), particle-particle
($\mathrm{pp}$) and hole-hole ($\mathrm{hh}$) intermediate-state
contributions are
\begin{widetext}
\begin{equation}
\begin{split}
W_S^{\mathrm{ph}} &= \frac{16V^4}{U}
\frac{1}{L^2} \sum_{k \leq k_F, k' > k_F}
\frac{8(\delta-\epsilon_k) (\delta-\epsilon_{k'}) (U+\epsilon_k-\epsilon_{k'})}
{(U-2\delta+2\epsilon_k)^2 (U+2\delta-2\epsilon_{k'})^2
(\epsilon_k-\epsilon_{k'})}, \\
W_T^{\mathrm{ph}} &= 32V^4
\frac{1}{L^2} \sum_{k \leq k_F, k' > k_F}
\frac{2U^2+5U(\epsilon_k-\epsilon_{k'})+4\left( \delta^2+\epsilon_k^2+\epsilon_{k'}^2
-\epsilon_k \epsilon_{k'} - \delta (\epsilon_k + \epsilon_{k'})\right)}
{(U-2\delta+2\epsilon_k)^2 (U+2\delta-2\epsilon_{k'})^2
(\epsilon_k-\epsilon_{k'})}, \\
W_S^{\mathrm{pp}} &= \frac{16V^4}{U}
\frac{1}{L^2} \sum_{k'_1 > k_F, k'_2 > k_F}
\frac{2(U-2\delta)(3U-2\delta)+8(U-\delta)(\epsilon_{k'_1}+\epsilon_{k'_2})+
8 \epsilon_{k'_1} \epsilon_{k'_2}}
{(U-2\delta+2\epsilon_{k'_1})^2 (U-2\delta+2\epsilon_{k'_2})^2},\\
W_S^{\mathrm{hh}} &= \frac{16V^4}{U}
\frac{1}{L^2} \sum_{k_1 \leq k_F, k_2 \leq k_F}
\frac{2(U+2\delta)(3U+2\delta)-8(U+\delta)(\epsilon_{k_1}+\epsilon_{k_2})+
8 \epsilon_{k_1} \epsilon_{k_2}}
{(U+2\delta-2\epsilon_{k_1})^2 (U+2\delta-2\epsilon_{k_2})^2}.
\end{split}
\end{equation}
\end{widetext}

\begin{figure}[htbp!]
\includegraphics[width=8cm,clip]{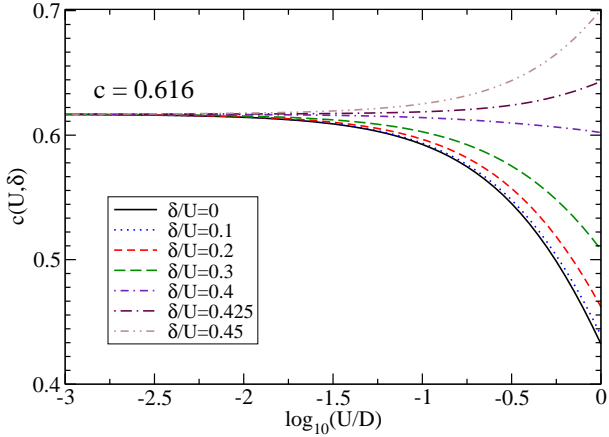}
\caption{The prefactor $c$ in the RKKY exchange constant 
$J_\mathrm{RKKY}=c 16 V^4/U$ for a flat band with $\rho_0=1/(2D)$
as a function of $U$ for a range of values of the impurity
energy level $\delta$.}
\label{fig_x}
\end{figure}

From these expressions we obtain $J_\mathrm{RKKY}=E_S-E_T$. In order to
evaluate the sums for a flat band with a constant density of states
$\rho_0=1/(2D)$ and the chemical potential $\mu=0$, we make formal
replacements $\frac{1}{L} \sum_{k'>k_F} = \frac{1}{L} \sum_{k'>0} \to 1/2
\int_0^1 dk'$ and $\frac{1}{L} \sum_{k \leq k_F} = \frac{1}{L} \sum_{k<=0}
\to 1/2 \int_{-1}^0 dk$. In Fig.~\ref{fig_x} we plot the prefactor $c$ in
the expression for the exchange constant $J_\mathrm{RKKY}=c 16 V^4/U$ as a
function of $U/D$. In the wide-band limit, i.e. for small $U/D$, $c$
approaches a constant value of $c=0.616$ irrespective of the value of
$\delta/U$. The dependence of $c$ on $\delta$ for $U/D \sim 1$
is due to the band-edge effects.

\begin{figure}[htbp!]
\includegraphics[width=8cm,clip]{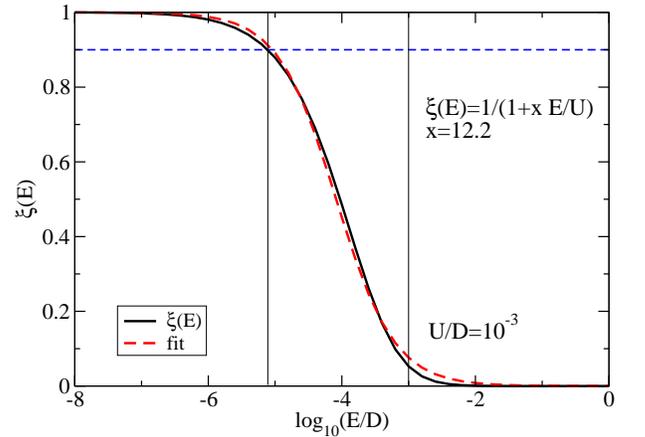}
\caption{Ratio $\xi(E)=J_\mathrm{RKKY}(E)/J_\mathrm{RKKY}^0$
of the running RKKY coupling constant at energy $E$ over
its value in the $E \to 0$ limit. The dashed line is an approximate
fit to a simple rational function $\xi(E)=1/(1+x E/U)$.}
\label{fig_y}
\end{figure}

To determine the temperature $T_J$ at which the RKKY interaction becomes
fully established, we calculate the cut-off dependent $J_\mathrm{RKKY}(E)$,
where $E$ is the low-energy cut-off for $k$ and $k'$ integrations, i.e. the
integrals over $k$ and $k'$ become $\int_{E}^1dk'$ and $\int_{-1}^{-E} dk$.
In Fig.~\ref{fig_y} we plot the ratio $\xi(E) =
J_\mathrm{RKKY}(E)/J_\mathrm{RKKY}^0$, where $J_\mathrm{RKKY}^0=
J_\mathrm{RKKY}(E\to 0)$. The ratio $\xi(E)$ reaches an (arbitrarily chosen)
value of $0.9$ at $E/U \sim 0.02$. This value of $E$ roughly defines $T_J$
below which the RKKY is fully developed. For small enough $V$ (i.e.
$\Gamma$), the value of $T_J$ is positioned between $T_1^*$ (free-orbital to
local-moment transition temperature) and $T_F^*$, the temperature of
ferromagnetic ordering of spins, given by $T_F^* = J_\mathrm{RKKY}^0 /
\beta$, where $\beta$ is a constant of the order one. For larger $V$,
however, $J_\mathrm{RKKY}(T)$ does not reach its limiting value at the
temperature where the spins start to order. In this case we obtain the
ordering temperature $T_F^*$ numerically as the solution of the implicit
equation
\begin{equation}
\label{selfconsist}
T_F^* = J_\mathrm{RKKY}(T_F^*)/\beta.
\end{equation}
An approximate fit to $\xi(E)$ in the wide-band limit is $\xi(E) = 1/(1+x
E/U)$ with $x=12.2$. We then obtain a solution for $T_F^*$ in closed form:
\begin{equation}
\begin{split}
T_F^* &= \frac{\sqrt{1+4x/\beta (J_\mathrm{RKKY}^0/U)}-1}{2x/U} \\
&\approx \frac{J^0_\mathrm{RKKY}}{\beta} 
\left( 1-\frac{x}{\beta} \frac{J_\mathrm{RKKY}^0}{U} + 
{\mathcal{O}} \left(\frac{J_\mathrm{RKKY}^0}{U}\right)^2 \right).
\end{split}
\end{equation}

\subsection{Six-fold symmetric $U_{12}=U$ case}

We study the splitting between the singlet, the triplet (same as above) and
the ``exciton'' states $\ket{I=0} = 1/\sqrt{2} \left(
\ket{\uparrow\downarrow,0} - \ket{0,\uparrow\downarrow} \right)$ and
$\ket{I=1} = 1/\sqrt{2} \left( \ket{\uparrow\downarrow,0} + \ket{0,
\uparrow\downarrow} \right)$. Second order corrections are all equal:
$E_S^{(2)} = E_T^{(2)} = E_{I=0}^{(2)} = E_{I=1}^{(2)} = - (S_1 + S_2)$
where $S_1$ and $S_2$ are the same as in the previously treated $U_{12}=0$
case. There is again no splitting to second order in $V$. The fourth order
corrections are
\begin{equation}
\begin{split}
E_S^{(4)} &= W_S^{\mathrm{ph}} + W_S^{\mathrm{pp}} + W_S^{\mathrm{hh}} \\
E_T^{(4)} &= W_T^{\mathrm{ph}} \\
E_{I=0}^{(4)} &= E_T^{(4)} \\
E_{I=1}^{(4)} &= E_S^{(4)}
\end{split}
\end{equation}
where
\begin{widetext}
\begin{equation}
\begin{split}
W_S^{\mathrm{ph}} &= 16 V^4
\frac{1}{L^2} \sum_{k \leq k_F, k' > k_F}
\frac{1}{\epsilon_k-\epsilon_{k'}}
\left(
\frac{1}{(U-2\delta+2\epsilon_k)^2}
+
\frac{1}{(U+2\delta-2\epsilon_{k'})^2}
\right), \\
W_T^{\mathrm{ph}} &= 32 V^4
\frac{1}{L^2} \sum_{k \leq k_F, k' > k_F}
\frac{3U^2 + 6U (\epsilon_k-\epsilon_{k'}) + 4\left(\delta^2+\epsilon_k^2
+\epsilon_{k'}^2-\epsilon_k \epsilon_{k'}-\delta (\epsilon_k+\epsilon_{k'})
\right)}
{(U-2\delta+2\epsilon_k)^2 (U+2\delta-2\epsilon_k)^2
(\epsilon_k - \epsilon_{k'})}, \\
W_S^{\mathrm{pp}} &= 16V^4
\frac{1}{L^2} \sum_{k'_1 > k_F, k'_2 > k_F}
\frac{2(U-2\delta+\epsilon_{k'_1}+\epsilon_{k'_2})^2}
{(U-2\delta+2\epsilon_{k'_1})^2(U-2\delta+2\epsilon_{k'_2})^2
(2U-2\delta+\epsilon_{k'_1}+\epsilon_{k'_2})}, \\
W_S^{\mathrm{hh}} &= 16V^4
\frac{1}{L^2} \sum_{k_1 \leq k_F, k_2 \leq k_F}
\frac{2(U+2\delta-\epsilon_{k_1}+\epsilon_{k_2})^2}
{(U+2\delta-2\epsilon_{k_1})^2 (U+2\delta-2\epsilon_{k_2})
(2U+2\delta-\epsilon_{k_1}+\epsilon_{k_2})}.
\end{split}
\end{equation}
\end{widetext}

The triplet is degenerate with the $I=0$ state, while the singlet and the
$I=1$ state are higher in energy, as determined by performing the
integrations (results not shown).

\section{Scaling equations to second order in $J$}
\label{scaling}

We consider an effective Hamiltonian of the form
\begin{equation}
\begin{split}
H &= \sum_{k\sigma} \epsilon_k c^\dag_{k\sigma} c_{k\sigma}
+ \sum_m E_m X_{mm} \\
&+ \sum_{mm',kk',\sigma\sigma'} J^{\sigma\sigma'}_{mm'}
X_{mm'} c^\dag_{k\sigma} c_{k'\sigma'},
\end{split}
\end{equation}
where $X_{mm'}=\ket{m}\bra{m'}$ are the Hubbard operators and
$J^{\sigma\sigma'}_{mm'}$ are generalized exchange constants.

We write \cite{hewson}
\begin{equation}
\begin{split}
& \Bigl[
H_{11}
+ H_{12} \left( E-H_{22} \right)^{-1} H_{21} \\
& + H_{10} \left( E-H_{00} \right)^{-1} H_{01}
\Bigr]\psi_1 = E \psi_1,
\end{split}
\end{equation}
where subspaces 2 corresponds to states with one electron in the upper
$|\delta D|$ edge of the conduction band, 0 corresponds to states with
one hole in the lower $|\delta D|$ edge of the band, and 1 corresponds
to states with no excitations in the edges that are being traced-over.
Furthermore, $H_{ij} = P_i H P_j$, where $P_i$ are projectors
to the corresponding subspaces $i$.

To second order, the coupling constant are changed by
\begin{equation}
\label{deltaJ}
\begin{split}
\delta J^{\sigma\sigma'}_{mm'} &= \rho_0 |\delta D|
\sum_{n\tau}
\frac{1}{E - D + \epsilon_k - E_n -H_0}
J^{\tau \sigma}_{nm} J^{\tau \sigma'}_{nm'}\\
&-\rho_0 |\delta D| \sum_{n\tau}
\frac{1}{E - D - \epsilon_{k'} - E_n - H_0}
J^{\sigma'\tau}_{nm} J^{\sigma\tau}_{nm'}.\\
\end{split}
\end{equation}

We apply these results to the effective low-temperature Kondo Hamiltonian
\begin{equation}
H_\mathrm{eff} = H_{\mathrm{band}} + J_1 \vc{s} \cdot \vc{S}_1 + J_2
\vc{s} \cdot \vc{S}_2 - J^{\mathrm{eff}}_\mathrm{RKKY} \left( \vc{S}_1 \cdot \vc{S}_2
- 1/4 \right).
\end{equation}
Introducing spin-1 operator $\vc{S}$ defined by the following Hubbard
operator expressions: $S_z = X_{\uparrow\uparrow} -
X_{\downarrow\downarrow}$, $S^+ = \sqrt{2} \left( X_{\uparrow0} +
X_{0\downarrow} \right)$ and $S^- = (S^+)^\dag$, we obtain
\begin{equation}
\begin{split}
H &= H_\mathrm{band} + {\tilde J} \vc{s} \cdot \vc{S}
+ J^\mathrm{eff}_\mathrm{RKKY} X_{SS} \\
&+\Delta \bigl( s_z (X_{0S} + X_{S0})
+ s^+ (X_{\downarrow S} - X_{S \uparrow}) \\
&+ s^- (X_{S \downarrow} - X_{\uparrow S}) \bigr),
\end{split}
\end{equation}
where index $S$ denotes the singlet state and we have
\begin{equation}
\label{Jinitial}
\begin{split}
{\tilde J} &= \frac{J_1 + J_2}{2} = J_0 (1+\alpha^2)/2, \\
\Delta  &= \frac{J_1 - J_2}{2} = J_0 (1-\alpha^2)/2.
\end{split}
\end{equation}

Equations \eqref{deltaJ} reduce to two equations for
$\tilde J$ and $\Delta$
\begin{equation}
\begin{split}
\delta {\tilde J} &= \rho_0 |\delta D| \left[
\frac{{\tilde J}^2}{D} + \frac{{\Delta}^2}{D+J^\mathrm{eff}_\mathrm{RKKY}}
\right], \\
\delta \Delta  &= -2\rho_0 \frac{|\delta D|}{D} \Delta {\tilde J}.
\end{split}
\end{equation}
from which ensue the following scaling equations
\begin{equation}
\begin{split}
\frac{d{\tilde J}}{dl} &= -\rho_0 {\tilde J}^2 - \rho_0 \frac{\Delta^2 D}
{D+J^\mathrm{eff}_\mathrm{RKKY}}, \\
\frac{d\Delta}{dl} &= - 2 \rho_0 \Delta {\tilde J},
\end{split}
\end{equation}
where $l=\log D$. The initial bandwidth $D$ is the effective bandwidth
$D_\mathrm{eff} = 0.182 U$ for the Anderson model and we take $\tilde
J(l=\log D_\mathrm{eff}) = \tilde J$ and $\Delta (l=\log
D_\mathrm{eff}) = \Delta$ with $\tilde J$ and $\Delta$ taken from
Eq.~\eqref{Jinitial}. We integrate the equations numerically until
$\tilde J$ starts to diverge. The corresponding cut-off $D$ defines
the Kondo temperature. The results are shown in Fig.~\ref{fig_r4}.
The scaling approach reproduces our NRG results very well.

\bibliography{vzporedne}

\end{document}